\documentclass{emulateapj}
\usepackage{apjfonts}
\usepackage{graphicx}
\usepackage{amsmath}
     \usepackage{dpfloat}

\usepackage{longtable}

\usepackage{color}

\def\kms{\ifmmode{\rm km\thinspace s^{-1}}\else km\thinspace s$^{-1}$\fi}

\shortauthors{Sanchis-Ojeda et al.~2014}
\shorttitle{The shortest-period planets}

\begin{document}

% ------------------------------------------------------------------------
% New commands
%
\def\ltsima{$\; \buildrel < \over \sim \;$}
\def\lsim{\lower.5ex\hbox{\ltsima}}
\def\gtsima{$\; \buildrel > \over \sim \;$}
\def\gsim{\lower.5ex\hbox{\gtsima}}
% -------------------------------------------------------------------------
%

\bibliographystyle{apj}

\title{
A Study of the Shortest-Period Planets Found With Kepler}

\author{
Roberto~Sanchis-Ojeda\altaffilmark{1},
Saul~Rappaport\altaffilmark{1},
Joshua~N.~Winn\altaffilmark{1}, 
Michael~C.~Kotson\altaffilmark{2},
Alan~Levine\altaffilmark{3}, 
Ileyk~El~Mellah\altaffilmark{4}
}

\altaffiltext{1}{Department of Physics, and Kavli Institute for
  Astrophysics and Space Research, Massachusetts Institute of
  Technology, Cambridge, MA 02139, USA; rsanchis86@gmail.com, sar@mit.edu, jwinn@mit.edu}

\altaffiltext{2}{Institute for Astronomy, University of Hawaii, 2680 Woodlawn Drive, Honolulu, HI 96822, USA}

\altaffiltext{3}{37-575 M.I.T.\ Kavli Institute for Astrophysics and Space Research, 70 Vassar St., Cambridge, MA, 02139}

\altaffiltext{4}{Laboratoire APC, Universit{\'e} Paris Diderot, France ; ileyk@apc.univ-paris7.fr}

 \slugcomment{Submitted to the {\it Astrophysical Journal}, 2014 February 4}

\begin{abstract}

  We present the results of a survey aimed at discovering and studying transiting planets with orbital periods shorter than one day (ultra--short-period, or USP, planets), using data from the {\em Kepler} spacecraft. We computed Fourier transforms of the photometric time series for all 200,000 target stars, and detected transit signals based on the presence of regularly spaced sharp peaks in the Fourier spectrum. We present a list of 106 USP candidates, of which 18 have not previously been described in the literature. In addition, among the objects we studied, there are 26 USP candidates that had been previously reported in the literature which do {\em not} pass our various tests. All 106 of our candidates have passed several standard tests to rule out false positives due to eclipsing stellar systems. A low false positive rate is also implied by the relatively high fraction of candidates for which more than one transiting planet signal was detected. By assuming these multi-transit candidates represent coplanar multi-planet systems, we are able to infer that the USP planets are  typically accompanied by other planets with periods in the range 1-50~days, in contrast with hot Jupiters which very rarely have companions in that same period range. Another clear pattern is that almost all USP planets are smaller than 2~$R_\oplus$, possibly because gas giants in very tight orbits would lose their atmospheres by photoevaporation when subject to extremely strong stellar irradiation. Based on our survey statistics, USP planets exist around approximately $(0.51\pm 0.07)\%$ of G-dwarf stars, and $(0.83\pm 0.18)\%$ of K-dwarf stars.

\end{abstract}

\keywords{planetary systems---planets and satellites: detection, atmospheres}

\section{Introduction}

The field of exoplanetary science rapidly accelerated after the discovery of hot Jupiters with orbital periods of a few days (Mayor et al.~1995; Marcy \& Butler 1996). More recently, another stimulus was provided by the discovery of terrestrial-sized planets with periods shorter than one day. These objects, which we will refer to as ultra-short period or USP planets, have many interesting properties. They are so close to their host stars that the geometric probability for transits can be as large as 40\%. The expected surface temperatures can reach thousands of kelvins, allowing the detection of thermal emission from the planets' surfaces (Rouan et al.~2011; Demory et al.~2012; Sanchis-Ojeda et al.~2013a). The induced stellar orbital velocities can be as high as a few m~s$^{-1}$, allowing the planet masses to be measured with current technology even for stars as faint as $V=12$ (Howard et al.\ 2013; Pepe et al.\ 2013). Among the best known USP planets are 55~Cnc~e (Dawson \& Fabrycky 2010; Winn et al.~2011; Demory et al.~2011a), CoRoT-7b (L{\'e}ger et al.~2009; Queloz et al.~2009), and Kepler-10b (Batalha et al.~2011).

The NASA {\em Kepler} space telescope (Borucki et al.~2010) monitored the brightness of about 200,000 stars for 4 years, long enough to observe thousands of transits of a typical USP planet. Along with Kepler-10b, some of the more prominent discoveries have been the innermost planets of Kepler-42 (Muirhead et al.\ 2012) and Kepler-32 (Fabrycky et al.~2012; Swift et al.~2013) and the system Kepler-70, where two very short period planets were inferred by means of the light reflected by their surfaces (Charpinet et al.\ 2011). However, since it was not clear that the official lists of {\em Kepler} USP planet candidates were complete, we and several other groups have performed independent searches. One object that emerged from our search was the Earth-sized planet Kepler-78b (Sanchis-Ojeda et al.~2013a), which has an orbital period of 8.5 hours and is currently the smallest exoplanet for which measurements of the mass and radius are both available (Howard et al.~2013; Pepe et al.~2013). Jackson et al.~(2013) performed an independent search for planets with periods $P<0.5$~day, finding several new candidates. More general surveys for {\em Kepler} planets have also found USP planets (Huang et al.~2013; Ofir \& Dreizler~2013). Particularly interesting is the discovery of KOI 1843.03 (Ofir \& Dreizler~2013), a planet with an orbital period of only 0.18~days or 4.25~hr. Rappaport et al.~(2013a) demonstrated that in order to survive tidal disruption, the composition of this Mars-sized planet must be dominated by iron as opposed to rock.

In this paper we describe a survey to detect USP planets using the entire {\em Kepler} dataset. Section \ref{sec:obs} describes the data that we utilized in our study.  Our Fourier-based transit search technique is explained in Section \ref{sec:search}, along with the steps that were used to winnow down thousands of candidates into a final list of 106 likely USP planets.  Section \ref{sec:trans} presents the properties of the candidates.  The issue of false positives within the USP list is examined in Section \ref{sec:multi}, with the conclusion that the false-positive probability is likely to be low. As a corollary we infer that most USP planets are accompanied by somewhat more distant planets. Section \ref{sec:occur} gives estimates for the occurrence rate of USP planets, and its dependence upon period, radius, and the type of host star. Finally, Section \ref{sec:disc} provides a summary of our findings and some remarks about the relevance of USP planets within the field of exoplanets.

\vspace{0.5cm}

\section{Observations}
\label{sec:obs}

\subsection{Kepler data}
\label{sec:kepdat}

To carry out an independent search for the shortest-period planets, we used the {\em Kepler} long-cadence time-series photometric data (30~min samples) obtained between quarters 0 and 16. A list was prepared of all $\approx$200,000 target stars for which photometry is available for at least one quarter, and the version 5.0 FITS files, which were available for all quarters, were downloaded from the STScI MAST website. We used the data that had been processed with the PDC-MAP algorithm (Stumpe et al.~2012; Smith et al.~2012), which is designed to remove many instrumental artifacts from the time series while preserving any astrophysical variability. We also made use of the time series of the measured $X$ and $Y$ coordinates of the stellar images (the ``centroids'') in order to test for false positives (see section~\ref{sec:prelim}). Finally, we used the information in the file headers to obtain estimates of the combined differential photometric precision (CDPP).  The CDPP is a statistic determined by the {\it Kepler} team's data analysis pipeline that is intended to represent the effective photometric noise level for detecting a transit of duration 6 hours (Christansen et al.\ 2012). We used this quantity in our analysis of survey completeness (see Section~\ref{sec:complete}).

\subsection{Stellar properties}

For estimates of basic stellar properties including not only radii, but also masses and effective temperatures, we relied upon the catalog of Huber et al.~(2014). This catalog is based on a compilation of photospheric properties derived from many different sources. Although it is not a homogeneous catalog, it likely provides the most accurate stellar parameters that are currently available. Stars for which only broadband photometry is available have radii that could be uncertain by up to $\sim$40\%, while stars for which spectroscopic or even astroseismic constraints are available have radius uncertainties as small as 10\%.

\begin{figure*}[ht]
\begin{center}
\leavevmode
\hbox{
\epsfxsize=5in
\epsffile{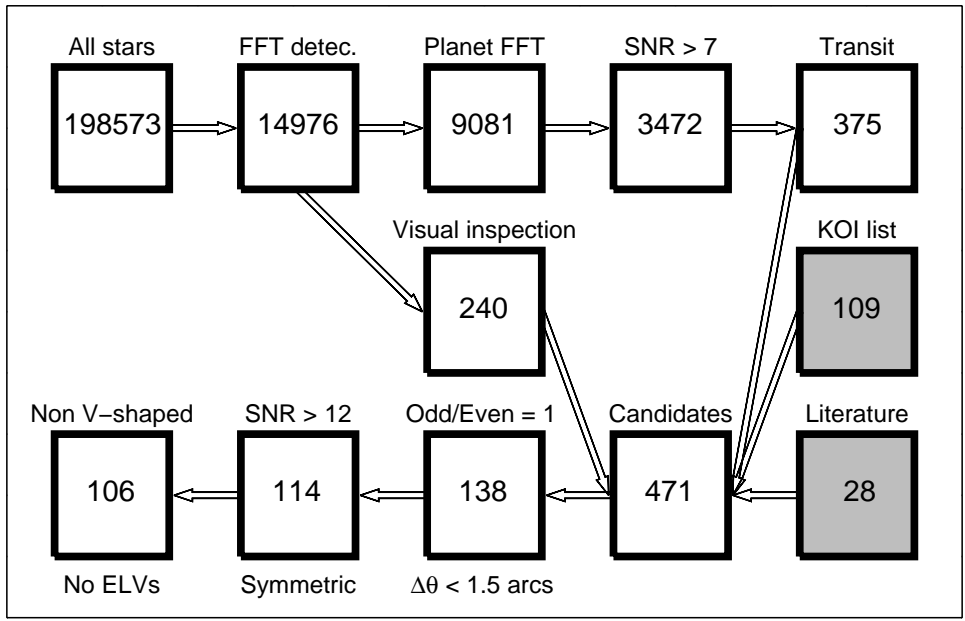}}
\end{center}
\vspace{-0.1in}
\caption{
{\bf Flow diagram for our USP planet search.}
The value in each box is the number of objects
that remained at each successive stage of the search,
beginning in the upper left corner with the 198,573 stars observed by {\em Kepler}.
The upper row represents the automated portion of the search, which, together with visual 
inspection of the folded light curves (fifth box) yielded 375 candidates.
This set of objects was combined with the objects that
were identified by visual inspection, and some that had already been designated
{\em Kepler} Objects of Interest (KOIs) or identified by other teams (gray boxes),
to yield a total of 471 distinct candidates that were studied in greater detail.
After applying several tests for false positives
caused by foreground and background binary stars (see Figure~\ref{fig:tests}),
and imposing a limiting signal-to-noise ratio of 12 in the folded light curve, 
we arrived at a final list of 106 USP candidates.}
\label{fig:diagram}
\vspace{0.1in}
\end{figure*}

\section{The search}
\label{sec:search}

\begin{figure*}[ht]
\begin{center}
\leavevmode
\hbox{
\epsfxsize=6in
\epsffile{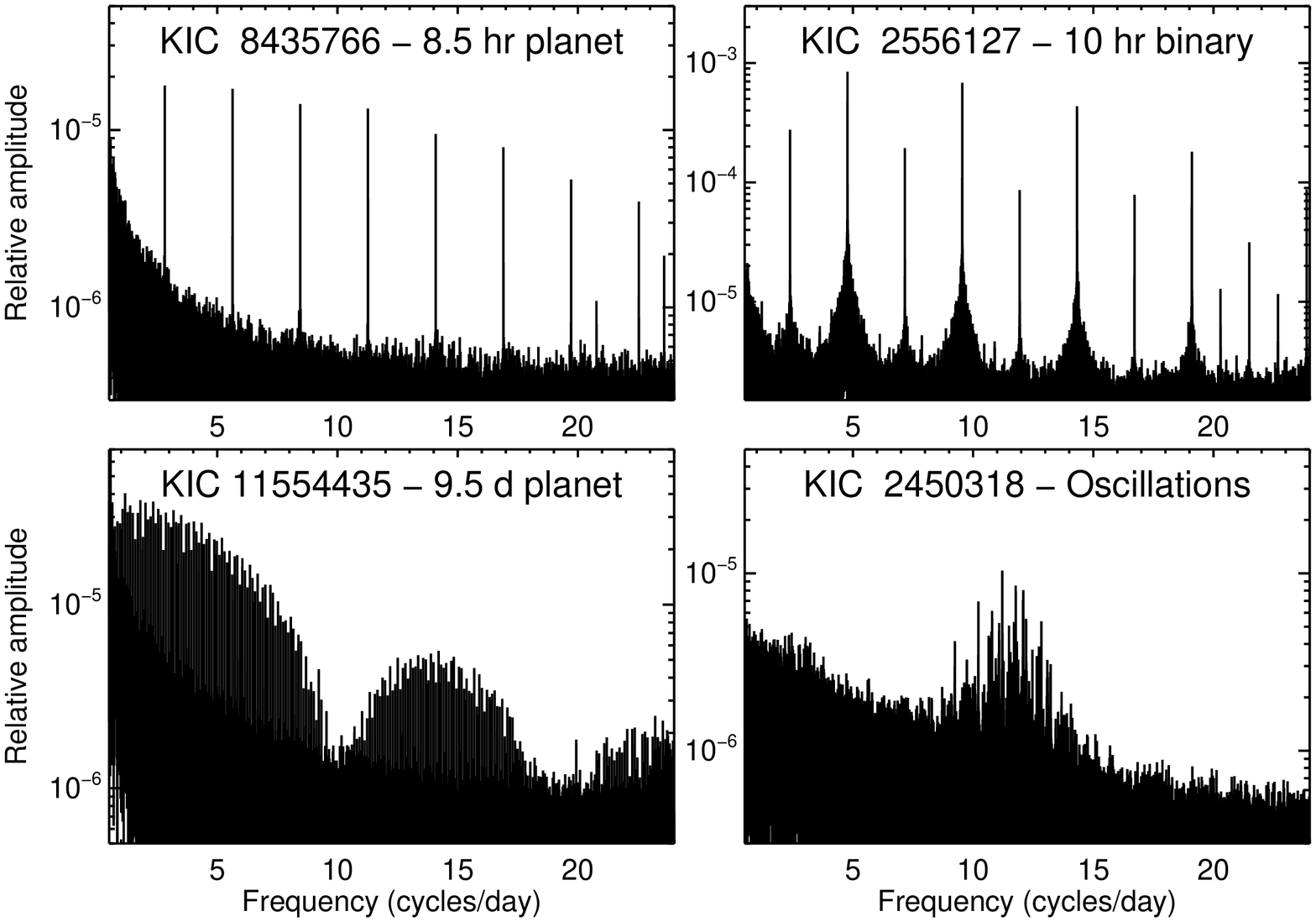}}
\end{center}
\vspace{-0.1in}
\caption{{\bf Illustrative FT amplitude spectra for different types of systems.}
{\it Upper left.}---A short-period planet (Kepler-78b; Sanchis-Ojeda et al.\ 2013a),
showing a series of harmonics with amplitudes that decrease monotonically and gradually
with frequency.
{\it Upper right.}---An eclipsing pair of stars, showing peaks that alternate in amplitude.
{\it Lower left.}---A long-period planet (Kepler-63b; Sanchis-Ojeda et al.\ 2013b),
showing very closely-spaced harmonics; the presence of long-period planets complicates
the Fourier-based detection of any short period planets.
{\it Lower right.}---Oscillations of a subgiant star. }
\label{fig:FFT}
\vspace{0.1in}
\end{figure*}

\subsection{The Fourier Transform Technique}

In our study of short-period planets, we elected to use a Fourier transform (``FT'') search. Since this is different from the standard algorithm for transit searching---the Box Least Squares (``BLS'') algorithm (Kov{\'a}cs et al.~2002)---it seems appropriate to provide some justification for our choice. The BLS is designed to have the greatest efficiency for transits with a duration that is short in comparison to the orbital period. The Fourier spectrum of an idealized transit light curve has a peak at the orbital period and a series of strong harmonics.  By using a matched filter, the BLS algorithm effectively sums all of the higher harmonics into a single detection statistic, which seems like a superior approach.

However, the standard BLS has a few drawbacks. One is that the BLS spectrum includes peaks at multiples of the orbital period {\it and} at multiples of the orbital frequency, thereby complicating attempts to ascertain the correct period. In addition, we have found that the standard BLS algorithm produces spurious signals at periods that are integer multiples of the {\em Kepler} sampling period of $\approx$0.02~day. These spurious peaks constitute a highly significant noise background in searches for planets with periods~$\lesssim$$0.5$~day. These spurious peaks can be partially suppressed by pre-whitening the data, i.e., attempting to remove the non-transit astrophysical variability prior to computing the BLS spectrum, but the introduction of such a step complicates the search.

A key advantage of the FT method is that FTs can be computed so quickly that it was practical to repeat the search of the entire database several times while the code was being developed (see for example the simulations by Kondratiev et al.~2009). Although the FT of a transit signal has power that is divided among several harmonics, the number of significant harmonics below the Nyquist limit declines as the orbital period is decreased, and the FT is therefore quite sensitive to the shortest-period planets. The ratio of transit duration to period, or duty cycle, varies as $P^{-2/3}$ and is as large as 20\% for USP planets, in which case the efficacy of the FT search is nearly equivalent to that of the BLS.  Furthermore, it is straightforward to detect a peak in the FT and its equally-spaced harmonics, either by means of an automated algorithm or by eye.  The absence of any subharmonics is a useful and important property of true planet transits as opposed to background blended binaries (which
often produce subharmonics due to the difference in depth between the primary and secondary eclipses). 

Regardless of the justification, it is often worthwhile to carry out searches with different techniques. Thus far, the independent searches of the {\it Kepler} database have utilized the BLS technique (Ofir \& Dreizler 2013; Huang et al.~2013; Petigura et al~2013; Jackson et al.~2013), and for this reason alone it seemed worthwhile to take a different approach. In the end, though, the proof of the effectiveness of an FT search lies in what is found, and in this work we demonstrate empirically that the FT is a powerful tool for finding short-period planets.

\subsection{Preliminary analysis of the candidates}
\label{sec:prelim}

Figure~\ref{fig:diagram} is a flow chart illustrating the numbers of the $\sim200,000$ {\em Kepler} stars that survived each stage of our search program. The PDC-MAP long-cadence time series data from each quarter was divided by the median flux of that quarter. We removed outliers with flux levels 50\% above the mean (likely due to cosmic rays), and all of the quarterly time series were stitched together into a single file. We further cleaned the data using a moving-mean filter with a length of 3 days. Using a filter to clean the data does not affect our ability to detect high frequency signals as long as the length of the window captures a few cycles of the target signal, and, at the same time, it increases the sensitivity to signals at intermediate frequencies, since it removes long-term trends that have the potential to increase the FT power at these frequencies. To prepare the data for the application of the Fast Fourier Transform (FFT), gaps in the time series were filled by repeating the flux of the data point immediately prior to the gap. The mean flux was then subtracted, giving a zero-mean, evenly-spaced time series. The FFT was then evaluated in the conventional manner all the way up to the Nyquist limit, i.e., the number of frequencies ($n_f$) in the transform is equal to half the number of data points.

We searched for the presence of at least one peak that is more than 4 times the local mean level in the Fourier amplitude spectrum (square root of the power spectrum), with a frequency higher than 1 cycle~day$^{-1}$. The local mean level was estimated using the closest $n_f/250$ frequencies, a number close to 120 for stars with 16 quarters of data. (This number is large enough to ensure that the interval used to estimate the local mean level is much wider than the FFT peaks, and that any possible contribution to the mean local level due to the peak is much smaller than the true local mean level.) In order for the target to be considered further, we also
required that the FT exhibit at least one additional harmonic that
stands out at least 3 times the local mean. For the selected objects, we increased the accuracy of the peak frequencies by calculating the FFT of a new flux series that was built from the original series by adding as many zeroes as needed to multiply the length of the array by a factor of 20 (a process known as ``frequency oversampling'').

Approximately 15,000 objects were selected by virtue of having at least one significant high-frequency peak and a harmonic. These objects underwent both visual and automated inspection.

First, all the Fourier spectra were examined by eye.  Those that showed peaks at several harmonics with slowly monotonically decreasing amplitudes, all of which were higher than 1~cycle~day$^{-1}$, were selected for further study (see Figure~\ref{fig:FFT} for some examples of the FT spectra). This process resulted in 240 objects worthy of attention, and was effective in quickly identifying the targets with the highest signal-to-noise ratios (SNRs) and periods shorter than 12-16 hours, such as Kepler-78b (Sanchis-Ojeda et al.\ 2013a).

Second, our automatic vetting codes were used to try to identify and reject unwanted sources such as long-period planets, eclipsing binaries, and pulsators. The initial step in this process was to find all significant FT peaks with an amplitude 4 or more times the local noise level. Next, all objects with more than 10 significant frequencies between 1 and 10 cycles~day$^{-1}$, a clear sign that the system is a pulsator or long-period planet and not a short-period transiting planet, were rejected. In other cases where frequencies lower than 1 cycle~day$^{-1}$ were detected and successfully identified as the true periodicity of the signal, the sources were also rejected. Application of these filters reduced the number of candidates from 15,000 to $\sim9,000$.

The surviving candidates then underwent a time-domain analysis. To remove the slow flux variations caused by starspots and stellar rotation, a moving-mean filter was applied to the flux series, with a width in time equal to the candidate orbital period. The data were then folded with that period, and the light curve was fitted with three models: a simple transit model, a sinusoid with the candidate period, and a sinusoid with twice the candidate period.  For a candidate to survive this step, the transit signal needed to be detected with SNR~$\geq 7$, and the transit model needed to provide a fit better than the fits of either of the sinusoidal models. This automated analysis reduced the number of objects from 9,000 to $\sim3,500$. 

The folded light curves were then inspected by eye. We found that in many cases the automated pipeline was still passing through some longer-period planets and pulsators. However, with only 3,500 objects, it was straightforward to reduce the list further through visual inspection of the folded light curves. We selected only those light curves with plausible transit-like features, including transit depths shallower than 5 to 10\%, to reduce the list of surviving candidates along this pathway to 375.

\begin{figure}[ht]
\begin{center}

\leavevmode
\hbox{
\epsfxsize=2in
\epsffile{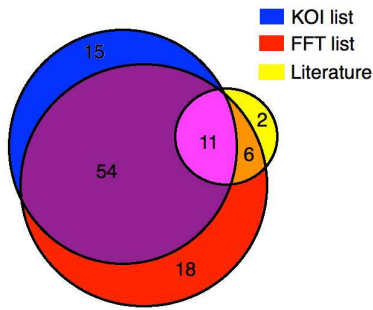}}
\end{center}
\vspace{-0.1in}
\caption{{\bf A Venn diagram describing the origin of the 106 USP candidates in our final list}. Our pipeline has been able to independently detect 89 out of these 106 candidates, with all missed planets having orbital periods longer than 12 hours. The literature contribution only counts those planets that were first detected by other groups (Huang et al.~2013; Ofir \& Dreizler~2013; Jackson et al.~2013) for a total of 19 USP planet candidates, since currently we have no knowledge about how many of the other 87 USP planet candidates they could have detected independently. 
}
\label{fig:venn}
\vspace{0.1in}
\end{figure}

We then combined the following lists: the 240 candidates from visual inspection of the FTs; the 375 candidates from the automated pipeline followed by visual inspection of the folded light curves; the 109 candidates with $P<1$~day in the KOI list (as of January 2014; Akeson et al.~2013); and the 28 candidates with $P<1$~day that emerged from the independent searches of Ofir \& Dreizler (2013), Huang et al.~(2013), and Jackson et al.~(2013). There was substantial overlap among these various lists (see Figure~\ref{fig:venn}). The result was a list of 471 individual candidates. The next step was to subject these candidates to selected standard {\em Kepler} tests for false positives (Batalha et al.\ 2010).

The first in this series of tests was the image centroid test which was used to reject those systems wherein the apparent motion of the stellar image during transits was large enough to rule out the {\em Kepler} target as the main source of the photometric variations. To carry out this test, the time series comprising the row ($X$) image centroid estimates and that comprising the column ($Y$) estimates were each processed with the moving-mean filter that was used for the flux time series.  Each filtered time series was folded with the candidate period. The degree of correlation between the flux deviations and the centroid deviations was then calculated by computing the values of $dY/df$ and $dX/df$, where $dX$ and $dY$ represent, respectively, the in versus out of transit change in row or column pixel position, and $df$ represents the in versus out of transit change in relative flux. These gradients were multiplied by the {\em Kepler} plate scale of 4 arcsec~pixel$^{-1}$ in order to estimate the location of the flux-variable source with respect to the center-of-light of the sources within the photometric aperture (Jenkins et al.~2010, Bryson et al.~2013). 

A significant centroid shift implies the presence of more than one star in the {\em Kepler} aperture, but does not necessarily imply that the object should be classified as a false positive; even if the intended {\em Kepler} target is indeed the source of the photometric variations, the steady light of other nearby stars would cause the centroid to move during transits.
Ideally, one would use the measured centroid shift and the known positions and mean fluxes of all the stars
within the {\em Kepler} aperture to pinpoint the variable star.
In the present case, though, the best available images (from the UKIRT {\em Kepler} field survey\footnote{\url{http://keplerscience.arc.nasa.gov/ToolsUKIRT.shtml}}) only allow us to differentiate fainter companion stars outside of a limiting separation that is target-dependent.
For example, a star 2 magnitudes fainter than the target star may be difficult to distinguish from the target star at separations less than an angle as small as $1\arcsec$ or as large as $3\arcsec$ (Wang et al.~2013). 

\begin{figure*}[ht]
\begin{center}

\leavevmode
\hbox{
\epsfxsize=5in
\epsffile{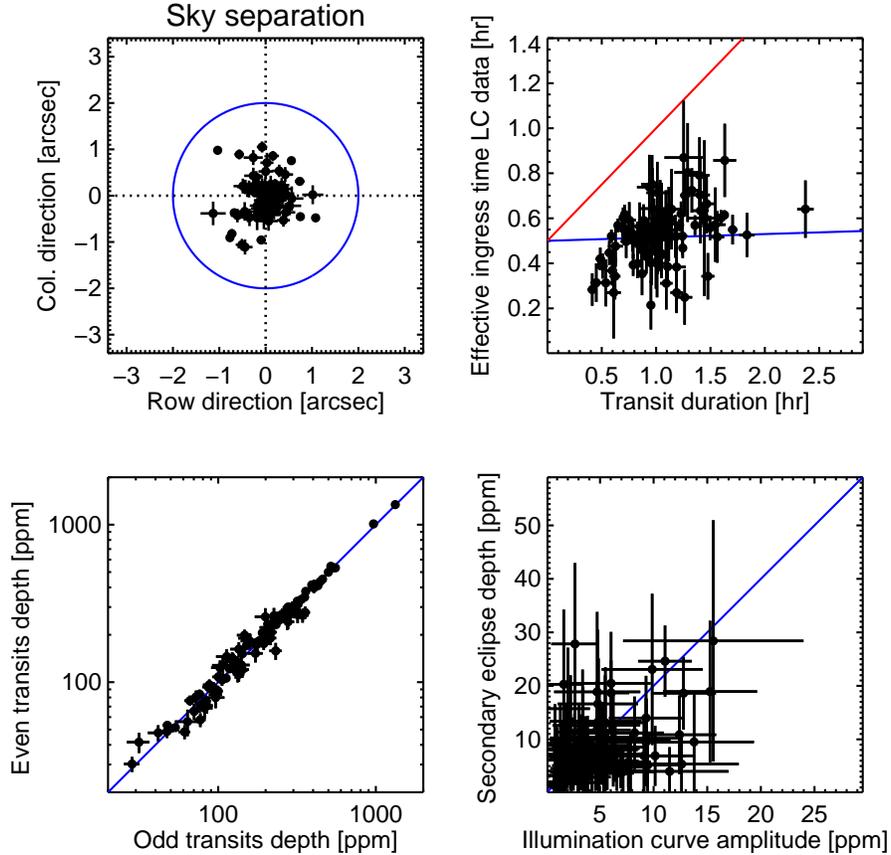}}
\end{center}
\vspace{-0.1in}
\caption{ {\bf The four main tests for false positives}.
Shown here are those 106 USP candidates in our final list, which have passed all our tests designed to rule out false positives.
{\it Upper left.}---Centroid test.  The measured change in position of the stellar images during transits was converted into a sky separation between the variable star and the center of light within the {\em Kepler} aperture.  The blue circle at 2 arcsec represents the approximate radius of confusion (see text) for all our objects.  {\it Upper right.}---Ingress/egress test. A comparison between the apparent ingress/egress time and total transit duration.  A small planet would produce a data point on the blue line, where the ingress time is approximately equal to 29.4~min (the sampling time of the observations), whereas a background eclipsing binary with comparably sized stars and a characteristically V-shaped light curve would produce a data point on the red line.  {\it Lower left.}---Odd/even test. The even and odd transit depths are statistically indistinguishable for all our candidates, constraining the possible background binaries to be close to identical stars.  {\it Lower right.}---For a true USP planet, any detected illumination curve should be accompanied by a secondary eclipse with a depth equal to twice the illumination amplitude (blue line). Otherwise, the out of eclipse variations could be attributed to ELVs from a background binary with twice the orbital period.  }
\label{fig:tests}
\vspace{0.1in}
\end{figure*}

Our procedure was to rely on the fact that very large centroid shifts almost always indicate
a false positive, since the {\em Kepler} target is, by construction, intended to be the dominant
source of light in the photometric aperture. The distribution of positions relative to the center of light among the objects under study may be reasonably modelled by the superposition of a uniform distribution out to distances beyond $40\arcsec$ and a Gaussian distribution centered at the center of light and having a width in each orthogonal coordinate characterized by a standard deviation close to $0.5\arcsec$. Based on this, we discarded those sources where the best-fit distance from the flux-variable source to the center-of-light exceeded $1.5\arcsec$ and a 3$\sigma$ lower bound exceeded $1\arcsec$.  This test assures that the ``radius of confusion'' (the maximum distance from the target star where a relevant background binary could be hiding) for almost all of the candidates is smaller than $2\arcsec$.  This is illustrated in the upper left panel of Figure~\ref{fig:tests}.  It was shown by Morton \& Johnson (2010) that by shrinking the radius of confusion to this level, the probability of false positives due to background binaries is reduced to of order 5\%.  This centroid test eliminated half of the remaining candidates, confirming that short-period eclipsing binaries in the background are an important source of false positives.

The second in this series of tests was a check for any statistically significant ($>$3$\sigma$) differences between the depths of the odd- and even-numbered transits. Such a difference would reveal the candidate to be an eclipsing binary with twice the nominal period (see lower left panel of Figure~\ref{fig:tests}).  Approximately 40\% of the objects were removed on this basis.

Third, there were five cases in which primary eclipses with a depth in the range 0.5-2\% were accompanied by secondary eclipses 5 to 10 times shallower, similar to what is expected for a high-albedo hot Jupiter in such a short period orbit. However, the orbits of all 5 objects were found to be synchronized with the rotation of their host stars, and a close inspection revealed that in all cases the secondary's brightness temperature (inferred from the secondary eclipses) was higher than what would be expected from a planet (assuming zero albedo and inefficient transfer of heat to the nightside), both of which are signs that these objects are probably low-mass stars rather than planets (see Demory et al.\ 2011b).

Finally, we eliminated every candidate with a transit signal-to-noise ratio less than~12, after judging that such detections are too weak to put meaningful constraints on false positive scenarios, and also out of concern that the pipeline might not be complete at low signal-to-noise ratios (as confirmed in Section~\ref{sec:complete}). This eliminated 22 more candidates. Within the remaining systems were two special cases, KIC~12557548 and KOI-2700 (Rappaport et al.\ 2012,~2013b), which have asymmetric transit profiles that have been interpreted as signs that these planets are emitting dusty comet-like tails.  Since the relationship between transit depth and planet radius is questionable for these objects, they were removed from our sample at this stage.

\section{Light curve analysis}
\label{sec:trans}

Once the number of USP candidates was reduced to 114, we
performed a more sophisticated light curve analysis. In particular, we wanted to make sure that systems with non-planetary transits or out-of-eclipse variations were completely removed. These tests, described below, also led to more reliable determinations of the transit parameters.

\subsection{Transit times and orbital period}
\label{subsec:timing}

Using a trial period obtained from the FT, we folded each light curve, binned it in phase, and fitted a simple trapezoidal model representing the convolution of the transit profile and the 30~min sampling function. The free parameters were the duration, depth, the ingress (or egress) time, and the time of the midpoint. 

With the best-fitting model in hand, the orbital period was recalculated according to the following procedure. For most of the candidates, individual transit events could not be detected with a sufficiently high SNR to obtain individual transit times.  Instead, we selected a sequence of time intervals each spanning many transits, and the data from each interval were used to construct a folded light curve with a decent SNR.  To decide how long the time intervals should be, we considered SNR$_q = $~SNR$/\sqrt{q}$, where SNR is the signal-to-noise ratio of the transit in the full-mission folded light curve and $q$ is the total number of quarters of data available. For the cases with SNR$_q < 10$, the intervals were chosen to span two quarters. For $10 <$~SNR$_q < 20$, the intervals spanned one quarter.  For $20 < $~SNR$_q < 30$, the intervals spanned one-third of a quarter (approximately one month), and finally for SNR$_q > 30$ the intervals spanned one-sixth of a quarter (approximately half a month). The trapezoidal model was fitted to each light curve to determine a mean epoch. For these fits, all parameters except the mean epoch were held fixed at the values determined for the full-mission folded light curve. Then, the collection of mean transit epochs was used to recalculate the orbital period. The formal uncertainty of each mean epoch was calculated using $\Delta \chi^2 = 1$, where $\chi^2$ was normalized to be equal to the number of degrees of freedom. We then fitted a linear function to the mean epochs. In this manner the final transit ephemeris was determined. We examined the residuals, and did not find any cases of significant transit-timing variations.  This finding is consistent with the prior work by Steffen \& Farr (2013) who noted that such short-period planets tend to avoid near mean-motion resonances with other planets.

\subsection{Transit and illumination curve analysis}
\label{subsec:timeseries}

\begin{figure*}[p]

\begin{center}

\leavevmode
\hbox{
\epsfxsize=7.0in
\epsffile{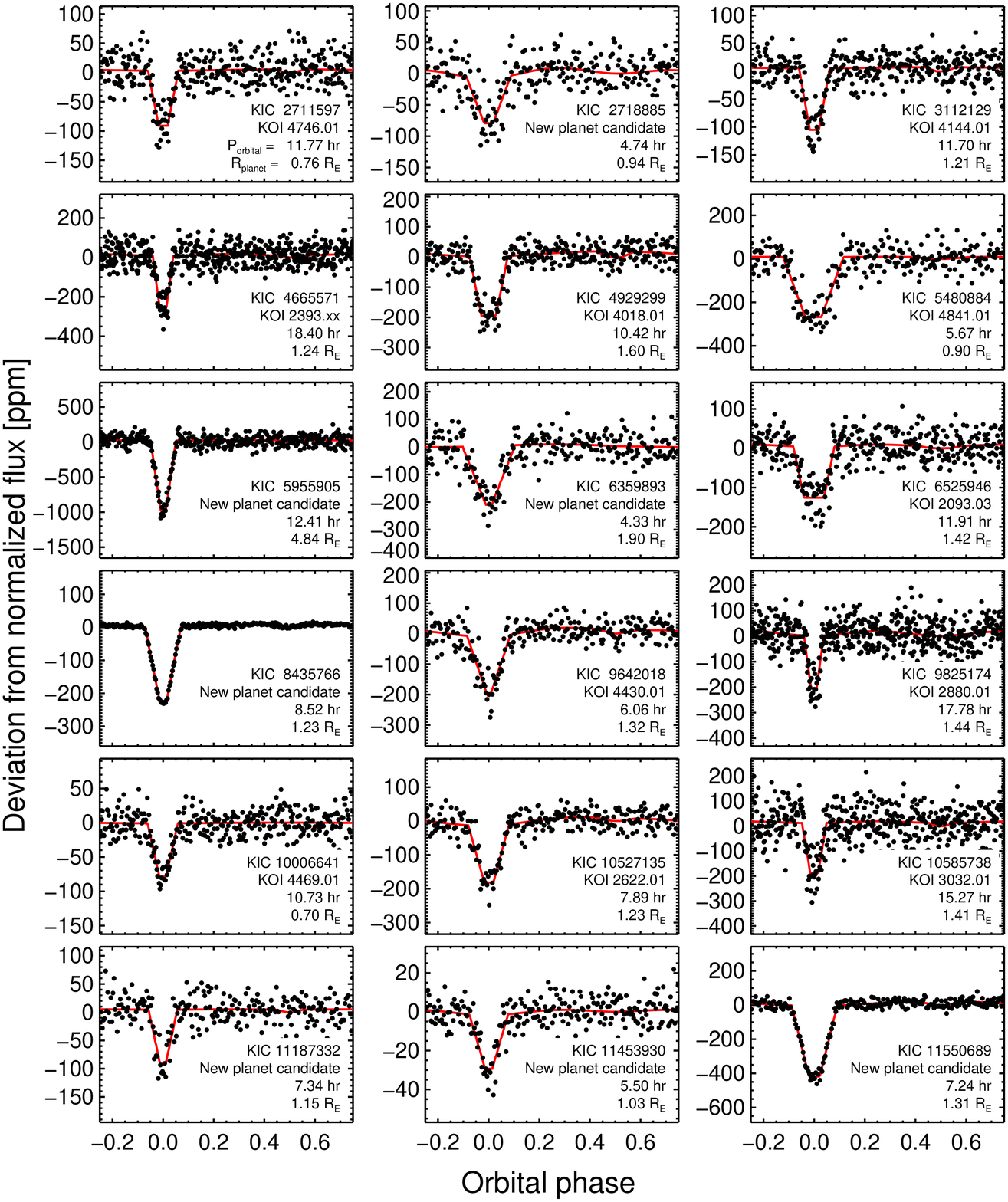}}
\end{center}
\vspace{-0.1in}
\caption{{\bf Transit light curves for new candidates}. These objects either
emerged from our search for the first time, or appear on the KOI list
as false positives but are considered by us to be viable candidates.
All fluxes are normalized to unity during secondary eclipse. The data have been
folded with the transit period and then averaged into 1 to 2-minute time bins (see text).
The red curve is the best-fitting transit model.}
\label{fig:new1}
\vspace{0.1in}

\end{figure*}

We then returned to the original time series and repeated the process of filtering out variability on timescales longer than the orbital period, including starspot-induced variations.  In this instance, the refined orbital period $P$ and a slightly different procedure were used. For each data point $f_0$ taken at time $t$, a linear function of time was fitted to the out-of-transit data points at times $t_j$ within $P/2$ of $t_0$.  Then $f_0(t)$ was replaced by $f_0(t) - f_{\rm fit}(t) +1$, where $f_{\rm fit}$ was the best-fitting linear function.  Illustrative filtered and folded light curves are shown in Figure~\ref{fig:new1}.

Further analysis was restricted to data from quarters 2-16, since quarters 0 and 1 were of shorter duration than the other quarters, and the data seem to have suffered more from instrumental artifacts. For each system, the data were folded with period $P$ and then binned to reduce the data volume and to enhance the statistics.  The bin duration was 2~min unless this resulted in fewer than 240 bins, in which case the bin duration was set to $P/240$ (as short as 1 minute for orbital periods of 4 hours). The trapezoidal model was fitted to this binned light curve. Due to the long cadence 30~min time averaging, the effective ingress duration of the transit of a typical short-period small planet (with a transit duration of one to two hours) will then be slightly longer than 30~min. However, in the case of a background binary, the ingress time can be much longer, up to half the duration of the entire event (see the upper right panel of Figure~\ref{fig:tests}). Once the best-fitting parameters were found, the allowed region in parameter space was defined with a Markov Chain Monte Carlo (MCMC) routine , which used the Gibbs sampler and the Metropolis-Hastings algorithm, and a likelihood proportional to exp($-\chi^2/2$) (see, e.g., Appendix A of Tegmark et al.~2004, or Holman et al.~2006). We found that five of the USP candidates had ingress times significantly longer than 45 minutes (with $>$3$\sigma$ confidence). These were removed from the candidate list.

With the list now reduced to 109 candidates, final transit parameters were determined. Our final model included a transit, an occultation, and orbital phase modulation. For the transit model we used an inverse boxcar (zero ingress time, and zero limb darkening) for computational efficiency. More realistic transit models are not justified given the relatively low SNR and the effects of convolution with the 30~min sampling function. The parameters were the midtransit time $t_0$, the transit depth $\delta_{\rm tran} \equiv (R_p/R_\star)^2$, and the transit duration.  The orbital period was held fixed, and a circular orbit was assumed.

The occultation (i.e., secondary eclipse) model $f_{\rm occ}(t)$ was also a boxcar
dip, centered at an orbital phase of 0.5, and with a total duration set equal to that of the transit model. The only free parameter was the occultation depth $\delta_{\rm occ}$.
The orbital phase modulations were modeled as sinusoids with phases and periods appropriate for ellipsoidal light variations (ELV), illumination effects (representing both reflected and reprocessed stellar radiation) and Doppler boosting (DB). Expressed in terms of orbital phase $\phi = (t-t_c)/P$,
these components are
\begin{equation}
- A_{\rm ELV}\cos(4\pi\phi)  - A_{\rm ill}\cos(2\pi\phi) + A_{\rm DB} \sin(2\pi\phi) ~~,
\label{eq:phase}
\end{equation}
respectively, where $t_c$ is the time of the transit center. The final parameter in the model is an overall flux multiplier, since only the relative flux values are significant and there is always an uncertainty associated with the normalization of the data. In the plots to follow, the normalizations of the model and the data were chosen to set the flux to unity during the occultations when only the star is visible. For comparison with the data, the model was evaluated every 30 seconds, and the resulting values were averaged in 29.4~min bins to match the time averaging of the {\it Kepler} data.  

Finally we determined the allowed ranges for the model parameters using a Monte Carlo Markov Chain algorithm. The errors in the flux data points in a given folded and binned light curve were set to be equal and defined by the condition $\chi^2=N_{\rm dof}$. No ELVs were found with amplitudes at or exceeding the $>$3$\sigma$ confidence level; hence, all of the candidates passed this test.  Secondary eclipses were detected in seven cases with amplitudes consistent with those of USP planets.  In three cases, an illumination component was detected while the corresponding secondary eclipses were not detected: $\delta_{\rm occ}/A_{\rm ill} < 2$ with 3$\sigma$ confidence. For one of them the DB component was also detected (the only system that did not past that test). Those three candidates were eliminated from the list because only a background binary with twice the nominal period could plausibly be responsible for the out-of-eclipse variability without producing a secondary eclipse (see the lower right panel of Figure~\ref{fig:tests}). We can also compare our results with some literature values. We obtained a secondary eclipse of $7.5 \pm 1.4$ ppm for Kepler-10b, a bit smaller than the value found by Fogtmann-Schulz et al.~(2014) of $9.9 \pm 1.0$ ppm, using only the SC data, but slightly larger than the original value reported by Batalha et al. (2010) of $5.8 \pm 2.5$. 

Based on the measured transit depth, and the stellar radius from Huber et al.~(2014), we calculated the implied planet radius. Based on our results for Kepler-78b (Sanchis-Ojeda et al.~2013a), we added a $20\%$ systematic uncertainty to the marginalized distribution of transit depths to account for the lack of limb darkening in our transit model. In almost all cases the uncertainty in the planet radius was still dominated by the uncertainty in the stellar radius.

One of the limitations of our transit analysis is that we could not readily obtain a value for the scaled semi-major axis ($a/R_\star$), which can ordinarily be obtained when fitting high-SNR transit light curves, and is needed for computing planet occurrence rates due to its role as the inverse geometric transit probability.  We can obtain a good estimate using the approximate relation:
\begin{equation}\label{eq:dur}
\frac{a}{R_\star} \approx \frac{P}{\pi T }\sqrt{1-b^2}
\end{equation}
where $P$ is the known orbital period, $T$ is the total duration of the transit, and $b$ is the impact parameter. For each of the values of $T$ obtained in the MCMC analysis, we evaluate the expression on the right-hand side using a value of $b$ drawn from a uniform distribution over the interval 0.0 to 0.9. The latter limit was not set to 1.0 because the transits in very high impact parameter cases are very short in duration and therefore very difficult to detect. Since the uncertainty in $a/R_\star$, estimated as the standard deviation of the final distribution, is large and dominated by the wide range of allowed $b$ values, the \textit{systematic} error induced by using equation~(\ref{eq:dur}) can be neglected.

Within our final list of 106 USP planet candidates, 8 candidates emerged uniquely from our search, including Kepler-78b (see Fig.~\ref{fig:venn} for a diagram describing the origin of all 106 candidates). Another 10 objects were flagged by the {\em Kepler} pipeline but were marked as false positives in the updated KOI list (Akeson et al.\ 2013), because, in at least some of the cases, the pipeline gave the wrong orbital period (the pipeline is not intended to work with periods below 12 hours).  The transit light curves of these 18 candidates are shown in Figure~\ref{fig:new1}, and the transit parameters are given in Table~\ref{tab:discovtab}. An examination of the KOI list shows that 27 KOI planet candidates with orbital periods shorter than 1 day were excluded in our analysis, either because of a low SNR (8 KOIs), or because those candidates did not pass our tests intended to exclude false positives (19 KOIs). Similarly, 7 planet candidates proposed by Ofir \& Dreizler (2013), Huang et al.~(2013), or Jackson et al.~(2013) did not appear in the KOI list, and also did not pass our tests. This information is summarized in Table~\ref{tab:removed}.

\begin{figure*}[p]
\begin{center}

\leavevmode
\hbox{
\epsfxsize=7.0in
\epsffile{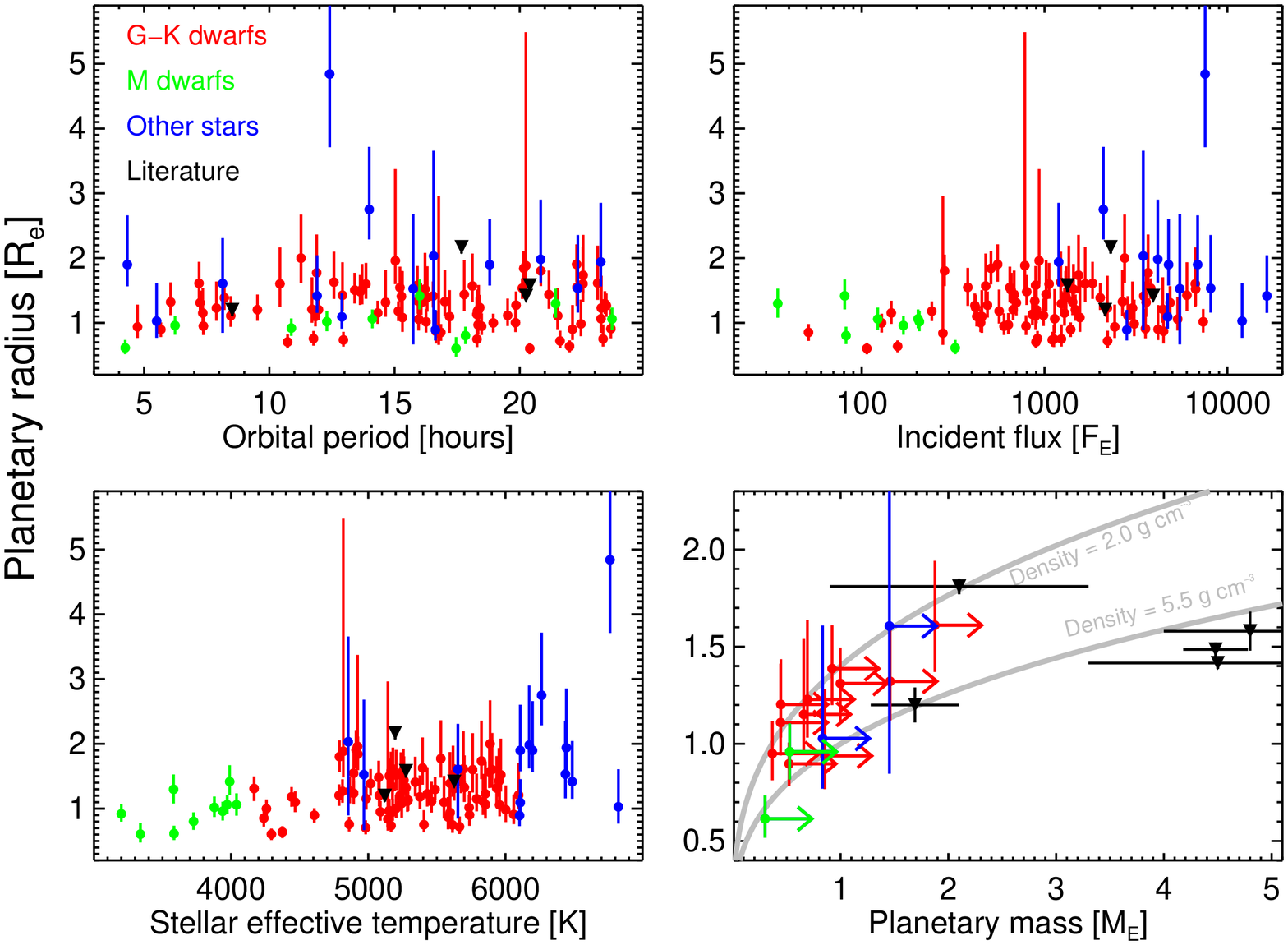}}
\end{center}
\vspace{-0.1in}
\caption{ {\bf Properties of the USP candidates.}
{\it Upper left.}---Planet radius vs.\ orbital period.
Red points are for candidates with
G and K dwarf host stars;
notably, all the candidates around such stars have a radius smaller than 2~$R_\oplus$.
Blue points are for candidates orbiting M dwarfs, F dwarfs,
or evolved stars. The two candidates with $R>2R_\oplus$ revolve around
KIC 5955905 and KIC 10281221. The triangles represent 55 CnC e (Winn et al.~2011; Demory et al.~2011a; Gillon et al. 2012), CoRoT-7b (L{\'e}ger et al.~2009, Queloz et al.~2009), Kepler-10b (Batalha et al.~2011), Kepler-78b (Sanchis-Ojeda et al.~2013a, Howard et al.~2013, Pepe et al.~2013).
{\it Upper right.}---Planet radius vs.\ incident bolometric flux, in units of the solar flux on Earth.
{\it Lower left.}---Planet radius vs.\ stellar effective temperature, showing no clear correlation except that the two planets larger than 2~$R_\oplus$
orbit stars hotter than 6250~K.
{\it Lower right.}---Constraints on mass and radius of the USP candidates (red and blue points) and other exoplanets drawn from the literature (black points).  For the USP candidates,
the mass constraint follows from the requirement that the orbit lie outside the Roche limit (Rappaport et al.~2013a).  Candidates with $P<10$~hr must have mean densities larger than 2~g~cm$^{-3}$; the shortest-period candidates must have mean densities exceeding that of Earth and are likely rocky. Additional black triangles represent Kepler-36b (Carter et al.~2012), and Kepler-11b (Lissauer et al.~2011, Lissauer et al.~2013), and 55 CnC e is too massive to appear in this panel.}
\label{fig:rvsp}
\vspace{0.1in}
\end{figure*}

\subsection{Planetary radius distribution}\label{sec:radius}

Figure~\ref{fig:rvsp} shows the radii and orbital periods of the 106 USP candidates.  In this figure we have identified the planets orbiting G and K dwarf stars ($4100$~K~$< T_{\rm eff}< 6100$~K, $4.0 < \log~g < 4.9$), which represent the subset of stars where most of the planets are detected. One striking feature of this group is the relative scarcity of planets larger than 2~$R_\oplus$.  This finding cannot be easily attributed to a selection effect, because larger planets are easier to detect than small planets.  This is in strong contrast to the planet population with periods in the range 2--100~days, for which planets with size 2-4~$R_\oplus$ (``sub-Neptunes'') are just as common as planets with radius 1-2~$R_\oplus$ (Howard et al.~2012; Fressin et al.~2013).

The explanation that comes immediately to mind is that gaseous planets are missing from the USP candidate list because such planets would have lost most of their gaseous envelopes to photoevaporation (Watson et al.~1981; Lammer et al.~2003; Baraffe et al.\ 2004; Murray-Clay et al.~2009; Valencia et al.\ 2009; Sanz-Forcada et al.~2011; Lopez, Fortney, \& Miller 2012; Lopez \& Fortney 2013a; Owen \& Wu 2013; Kurokawa \& Kaltenegger 2013).  Indeed, the levels of irradiation suffered by these planets (see Figure~\ref{fig:rvsp}) are high enough to remove the entire envelope from a gaseous planet for a wide range of mass loss efficiencies and core masses (Lopez \& Fortney 2013a). If that is the case, then the USPs may be a combination of rocky planets and formerly gaseous planets that lost their atmospheres. In fact, Lopez \& Fortney (2013a) predicted that planets smaller than Neptunes with incident fluxes larger than a hundred times the Earth's incident flux would end their lives mostly as rocky planets smaller than 2~$R_\oplus$.

In the observed radius distribution, 95\% of the candidates have a radius smaller than 1.9~$R_\oplus$.  If we assume that the USP candidate list contains only rocky planets (because gas would have been evaporated), and that it represents a fair sampling of the full range of possible sizes of rocky planets, then this finding can be interpreted as a direct measurement of the maximum possible size of a rocky planet. The value of 1.9~$R_\oplus$ is in agreement with a limiting size of 1.75~$R_\oplus$ quoted by Lopez \& Fortney (2013b) based on the complexity of forming planets larger than this limit with no significant H/He envelopes. Weiss \& Marcy (2013) also note that planets smaller than 1.5~$R_\oplus$ seem to have, on average, densities similar to that of Earth, which could be a sign that they are mostly rocky. It would be interesting to firm up our empirical determination of the limiting size of rocky planets after the estimates of the stellar radii have been improved using spectroscopy or other means.

There are also two USP candidates with implied planet radii larger than 2~$R_\oplus$. They both belong to the group of six planet candidates that orbit stars hotter than 6250~K. The probability of this occurring by chance is only 0.3\%, and it is intriguing that these hotter stars are distinguished from the cooler stars by the lack of a convective envelope (Pinsonneault et al.~2001). Perhaps the weaker tidal friction associated with the lack of a convective envelope has allowed these planets to survive tidal decay, and, furthermore, their core masses could be large enough that some of their gaseous envelopes have been retained (see Owen \& Wu~2013 and references therein).

For the shortest-period planets, meaningful lower limits on the mean densities can be established from the mere requirement that they orbit outside of their Roche limiting distances. We have calculated these limiting densities for all the planet candidates with orbital periods shorter than 10 hours orbiting G and K dwarfs, using the simplified expression (Rappaport et al.\ 2013a)
\begin{equation}
\rho_{\rm p} [{\rm g~cm}^{-3}] \geq \left(\frac{11.3~{\rm hr}}{P_{\rm orb}}\right)^2
\end{equation}
which is based on the assumption that the planet's central density
is no more than twice the mean density.
From this lower limit on the density, and the best fit estimated planet radius,
we were able to calculate a lower limit on the planet mass.
The results are shown in Figure~\ref{fig:rvsp}.
Including the USP sample, it is now possible to
place constraints on the mean density of about 10 additional terrestrial-sized planets.
For a few of them, namely those with orbital periods shorter than 5-6 hours, the minimum mean density is large enough that the planets are likely to be rocky. The caveats here are that we have not established individual false positive rates for most of these planets; and without spectroscopically determined properties for the host star, the uncertainties in the planet properties are large. 

\section{Completeness and false positive rate}
\label{sec:multi}

We will soon turn to the calculation of the ``occurrence rate'' of USP planets, defined as the probability that a given star will have such a planet.  Before calculating the occurrence rate, there are two important issues to address.   The first issue is completeness: how many transiting USP planets could have been missed in our search?   The second issue is that of false positives: how many of the 106 USP candidates are likely to be true planets as opposed to diluted eclipsing binaries or other systems that somehow mimic a planetary signal and evaded our various tests?

\subsection{Completeness}
\label{sec:complete}

Our search procedure was designed mainly to detect a sizable sample of new USP candidates, and in this respect it was successful, having yielded 18 new candidates.  However, the procedure was complicated enough that it is not straightforward to perform quantitative tests for completeness.  At least there are good signs that the completeness is high, based on a comparison between the yield of our pipeline, and the overlap with the various other searches that have been conducted with the {\em Kepler} data (see Figure~\ref{fig:venn}).

We have attempted to measure the completeness of the automated portion of our search by injecting simulated transits into real {\it Kepler} light curves and then determining whether they are detected by our pipeline analysis.  A similar inject-and-recover test for longer-period planets was recently performed by Petigura et al.\ (2013). A more realistic simulation in which the signatures of background binaries and other false positive sources, as well as planet transits, are injected into pixel level data, was beyond the scope of our study. 

To start, we selected a set of $105,300$ {\it Kepler} target stars consisting of G-dwarfs and K-dwarfs with $m_{\rm Kep}<16$. A star was then chosen at random from this set, an orbital period was chosen from a distribution with equal probabilities in equal logarithmic intervals over the range 4 hours to 24 hours, a planet radius was chosen from the same type of distribution, but over the size range 0.84~$R_\oplus$ to 4~$R_\oplus$, and a transit impact parameter was chosen at random from the interval 0 to 1. Given the period and the stellar properties, the orbital radius could be estimated from the relation $a/R_\star = 3\pi/(GP^2 \rho_\star^3)$. The duration of each transit was then calculated using Eqn.~(\ref{eq:dur}), and the planet radius and stellar radius were used to compute the transit depth $(R_p/R_\star)^2$. Simulated transit signals were then constructed using the trapezoidal model described in section~\ref{subsec:timeseries}, which takes into account the 30-minute cadence of the observations, and injected into the mean-normalized PDC-MAP light curve for the selected star.

This procedure, starting with the random selection of a star, was repeated, on average, twice for each of the 105,000 stars in the list, and all of the resulting flux time series were processed with the part of the automated pipeline corresponding to the first four boxes in Figure~\ref{fig:diagram} as used in our actual planet search. In brief, a 3-day moving-mean filter was applied to the light curve, and the FT was calculated.  The planet was considered further if the correct frequency or an integral multiple of the correct frequency was detected in the Fourier spectrum. Given a detection, a filtered folded light curve was constructed, using the detected period, and the planet was rejected if the SNR of the transit dip was lower than 7 or if a sinusoid yielded a better fit than a transit profile.  The final step was to remove those objects for which the transit signals have an SNR lower than 12 at the real orbital period (which may differ from the SNR at the detected orbital period).

We did not attempt to simulate the visual-inspection portions of the pipeline, as this would be prohibitively time-consuming (this issue is discussed further at the beginning of Section~\ref{sec:occur}). The false positive tests also were not performed, since the transit shapes should be similar to the injected transit shapes. 

\begin{figure*}[ht]
\begin{center}

\leavevmode
\hbox{
\epsfxsize=6.5in
\epsffile{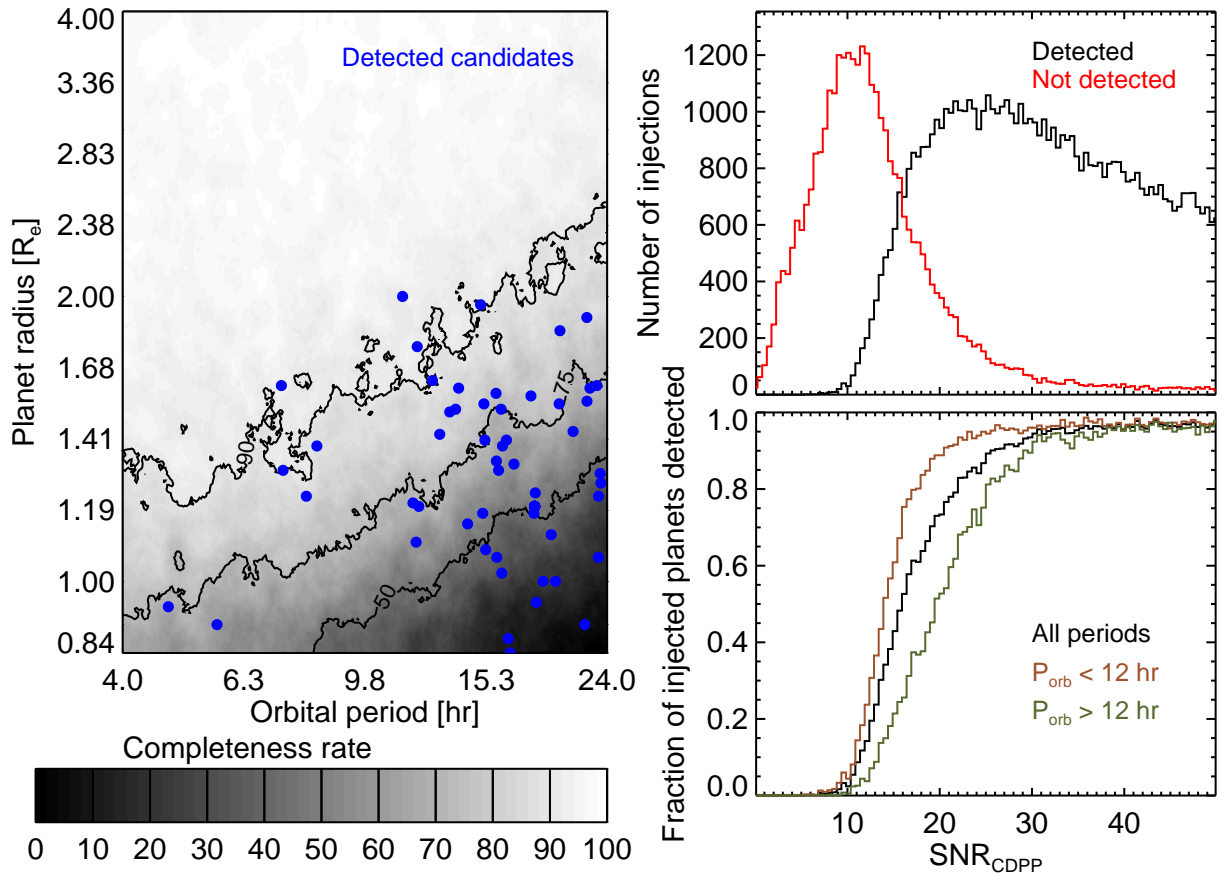}}
\end{center}
\vspace{-0.1in}
\caption{ {\bf Completeness of the automated FT pipeline,}
for detecting USP planets orbiting G and K dwarfs with $m_{\rm Kep}<16$.
{\it Left.}---The percentage of the artificially injected planet signals that were recovered by our automated pipeline. Three contours are plotted as thick black lines, representing completenesses of 90, 75 or 50\%. Blue dots represent the actual 69 USP planets detected by the same pipeline transiting the selected sample of {\it Kepler} stars. 
{\it Upper Right.}---Histogram showing the distribution of the SNR$_{\rm CDPP}$ (see text) of the detected and non-detected planet candidates. 
{\it Lower Right.}---The distributions shown in the upper right panel were used to obtain the fractions of planets (in black) detected at different values of SNR$_{\rm CDPP}$. The detection efficiency is near 100\% for SNR$_{\rm CDPP}$ > 30, and drops almost to zero for SNR$_{\rm CDPP} \approx 10$. The efficiency seems to decrease with the orbital period as expected, as shown by the differences between the efficiencies for planets with orbital periods shorter than 12 hours (in brown) and between 12 and 24 hours (in green). 
}
\label{fig:inject}
\vspace{0.1in}
\end{figure*}

The results of these numerical experiments are shown in Figure~\ref{fig:inject}. The left portion of the figure shows the dependence of the completeness on period and planet radius. Evidently the pipeline was able to detect nearly all planets larger than 2~$R_\oplus$.  The efficiency is also seen to generally rise at shorter periods.  This may be simply explained, at least in part, by the fact that more transits occur for shorter-period planets over the duration of the observations, which usually results in a higher total SNR than that of longer-period planets of the same size and with the same type of host star.

The right side of Figure~\ref{fig:inject} gives information relating to the efficiency of the pipeline as a function of the ``fiducial SNR'' of the transit signals. We calculated the fiducial SNR of the folded transit light curve based on the average over the available quarters of the CDPP noise levels that were reported in the headers of the FITS files (see section~\ref{sec:kepdat}; Howard et al.~2012) according to 
\begin{equation}
SNR_{\rm CDPP} \equiv \frac{\delta_{\rm tran}}{\sigma_{\rm CDPP, 6~hr}}\sqrt{\frac{T[{\rm hr}] (84.4\,q\, {\rm [days]})}{(6\,{\rm hr})~P_{\rm orb}{\rm [days]}}}
\end{equation}
where $q$ is the number of quarters of data, the factor 84.4 represents the average number of effective days of observations per quarter and the factor of 6~hr is introduced because the CDPP is reported for a 6~hr timescale. This definition of fiducial SNR is convenient because the level of noise in the light curve depends on the filter used, but the CDPP noise is easily accessible through the {\em Kepler} FITS files. As seen in the upper right panel of Figure~\ref{fig:inject}, our original $SNR > 12$ cut (see section~\ref{sec:prelim}) does not perfectly translate into $SNR_{\rm CDPP} > 12$. There is a small fraction (a few hundred) of simulated planets that were detected with $SNR_{\rm CDPP} < 12$, and there were also some planets with $SNR_{\rm CDPP} > 12$ that were missed by the pipeline. The efficiency increases from nearly zero at $SNR_{\rm CDPP} = 10$ (due to our $SNR > 12$ cut) to for example $90\%$ at $SNR_{\rm CDPP} = 25$ (also in Figure~\ref{fig:inject}).

This ``ramp'' in detection efficiency is reminiscent of the work of Fressin et al.~(2013), who proposed that the detection efficiency of the {\it Kepler} transit search pipeline rises as a nearly linear function of $SNR_{\rm CDPP}$ from essentially zero at $SNR_{\rm CDPP} = 7$ to approximately $100\%$ at $SNR_{\rm CDPP} = 16$. As one might have expected, our FT-based pipeline seems to be somewhat less efficient at detecting low-SNR planets than the {\em Kepler} pipeline.  The likely reason is that the significance of the FT-based detection is spread over numerous harmonics, and since we used only two harmonics to detect signals, some of the lowest-SNR planets are not detectable. Evidence for this can be seen in the lower right panel of Figure~\ref{fig:inject}, where the detection efficiency has been calculated for orbital periods shorter and longer than 12 hours. The pipeline detects many more low-SNR planets with short orbital periods, because their signals are spread over fewer harmonics. This may be the reason why most of our newly detected planet candidates have short orbital periods, while the candidates missed by our search and discovered by the {\em Kepler} pipeline tend to have longer orbital periods.

\subsection{Low false positive rate and high multiplicity}

\begin{figure*}[ht]
\begin{center}

\leavevmode
\hbox{
\epsfxsize=6.5in
\epsffile{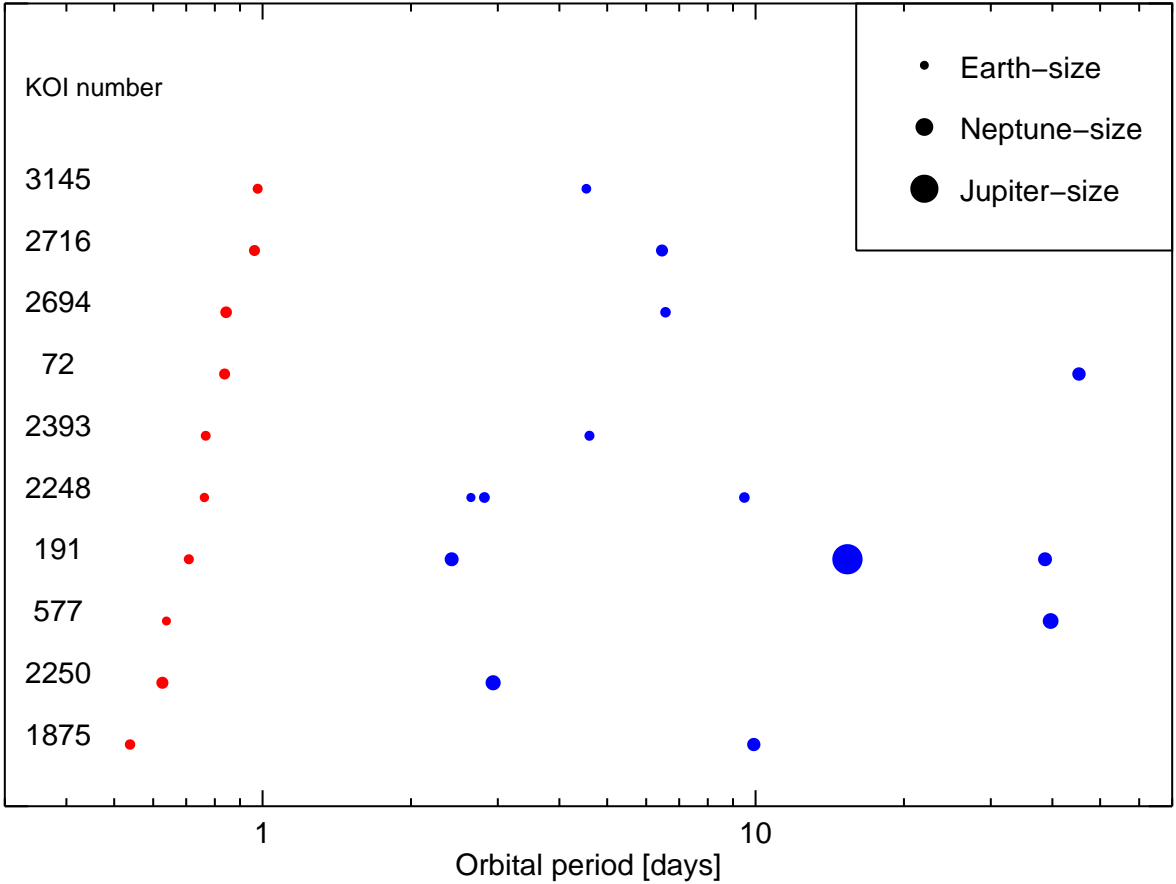}}
\end{center}
\vspace{-0.1in}
\caption{ {\bf Multitransiting systems with USP planets.}
Depicted are the periods and planet sizes for the subset of 10 USP candidates
detected with our pipeline with G and K-type host stars and at least one other transit candidate with $P=1$-50~days.
The radius of the dot is proportional to the square root of the planet radius.
The presence of multiple candidates causes these 10 USP candidates to have a very low false positive probability (Lissauer et al.\ 2012).
}
\label{fig:lp}
\vspace{0.1in}
\end{figure*}

We now investigate the issue of false positives.  One approach would be to pursue the validation or confirmation of all 106 objects, most definitively by direct mass measurement (as was done for Kepler-78; Howard et al.\ 2013; Pepe et al.\ 2013) or, somewhat less definitively, by the lack of ELVs and by statistical arguments (as was done for KOI-1843.03; Rappaport et al.\ 2013).  We have found that a simpler argument is available for the USP candidates, based on the empirically high probability that USP planets are found in compact, coplanar, multi-planet systems.

It has already been shown that the false positive rate for a transit candidate that is found in association with other transit candidates of different periods around the same {\em Kepler} target is much lower than the false positive rate of a transit candidate that is found in isolation (Lissauer et al.\ 2012; Lissauer et al.\ 2014). The reason is that background binary stars (an important source of false positives) are expected to be spread randomly over the {\em Kepler} field of view, whereas the planet candidates are found to be clustered around a relatively smaller sample of stars. This same line of argument can be extended to the USP candidates.  To be specific, in what follows we will restrict our attention to candidates detected with our pipeline and with G and K-type hosts.  For convenience we will also define a ``short-period planet'' (SP planet) as a planet with a period in the range 1-50~days, as opposed to USP planets with $P<1$~day. 

There are 10 stars hosting USP candidates in our sample of G and K stars that also host at least one transiting SP candidate. The orbital periods and planet sizes for these 10 systems are displayed in Figure~\ref{fig:lp}. These planet candidates are likely to be genuine planets, based on the statistical arguments by Lissauer et al.\ (2012). In order for this argument to work, false positives with orbital periods shorter than one day should be found less frequently orbiting multiplanet systems (Lissauer et al.~2014).  In particular, if we focus only on the active KOI list (Akeson et al.\ 2013), we evaluate the fraction of false positives detected for G and K-type hosts with $P<1$~day as being 6 times lower for stars that have at least one active SP planet candidate than for those that do not. Since all of our 10 USP candidates accompanied by transiting SP candidates have also passed conservative false positive tests, we feel justified in assuming that they have a high probability of being genuine planets, which we will take to be $\simeq 100\%$. Next we show that the existence of these 10 systems {\it implies} the existence of a larger number of transiting USP planets which have {\it non-transiting} SP planets. This is because the geometric probability of transits decreases with orbital distance, and is lower for the SP planet than for the USP planet.

The transit probability is given by $R_*/a \propto P^{-2/3}$ for stars of the same size.  We can derive an effective transit probability for a {\em range} of orbital periods by integrating $P^{-2/3}$ against the period distribution of planets, $f(P)$.  For the latter we utilize Eqn.~(8) of Howard et al.~(2012):
\begin{eqnarray}
f(P) \propto P^{\beta -1} \left(1-e^{-(P/P_0)^\gamma}\right)
\end{eqnarray}
where, for their case which includes all planet sizes, $P_0 \simeq 4.8$ days, $\beta \simeq 0.52$ and $\gamma \simeq 2.4$ (Howard et al.~2012).  We then compute the effective transit probability, $\mathcal{P}$
\begin{eqnarray}
\mathcal{P}(P_1 \rightarrow P_2) \propto \left(\int_{P_1}^{P_2} f(P)\, P^{-2/3} \,dP \middle/ \int_{P_1}^{P_2} f(P) dP\right)
\end{eqnarray}
When we carry this integral out numerically for the ranges: $P = 1/6 \rightarrow 1$ day and $P = 1 \rightarrow 50$ days, we find:
\begin{eqnarray}
\frac{\mathcal{P}(1/6 \rightarrow 1)}{\mathcal{P}(1 \rightarrow 50)} \simeq 6.9
\end{eqnarray}
Indicating the effective transit probability is $\sim$ 7 times higher in the USP range than in the SP range.
If the distribution of SP planets is unaffected by the presence of a USP planet, this then implies that if we find $10\pm \sqrt{10}$ USP planets with transiting SP planets, there should be a total of $69\pm 22$
USP planets with SP planet companions.\footnote{Here, in computing the uncertainties,
we have simply assumed that the number of detections is subject
to the usual counting statistics.} 
We have found 69 USP candidates orbiting G and K dwarfs, which is fully consistent with the idea that essentially all USP planets are in systems with other SP planets, whether they are seen transiting or not.  This also suggests that the number of false positives among the 69 USP planet candidates is likely to be quite low.

We do acknowledge that part of the above argument is somewhat circular since we have extrapolated the expression for $f(P)$ from Howard et al.~(2012), which was only derived for planets with $P > 1$ day and all sizes, into the USP range and for small planets.  We do, however, find numerically that the result is rather insensitive to the choice of $\gamma$ at least over the range $2 < \gamma < 5$, $\gamma$ being the parameter which largely dictates the slope in the short-period range.  Moreover, it turns out that the Howard et al.~(2012) extraplolated slope for $f(P)$ is essentially the same for their smallest planets as compared with all planets.  Thus, at a minimum, this very plausible period distribution function provides an interesting self-consistency check.  

In summary, it is very clear that the 10 USP candidates with transiting SP companions imply the existence of a much larger number of USP candidates with non-transiting SP planets -- and we indeed find them. 

Interestingly, Schlaufman et al.\ (2010) predicted that a large family of terrestrial planets would be found with {\em Kepler} with periods shorter than 1 day, and that they would be accompanied by other planets in their systems.
This was based on the fact that the dynamical interactions between planets
can bring a close-in planet even closer to its star, where tidal dissipation
can further shrink the orbit. However, it is unclear if planets with orbital periods as long as 50 days could have had any dynamical influence on the USP, particularly since these planets are also quite small (see Figure~\ref{fig:lp}). What is clear, though, is that the USPs are very different from hot Jupiters, which have been found to have a very low probability of having additional companions with periods $<$50~days (Steffen et al.~2012).

\section{Occurrence rate}
\label{sec:occur}

\begin{figure*}[ht]
\begin{center}

\leavevmode
\hbox{
\epsfxsize=6.5in
\epsffile{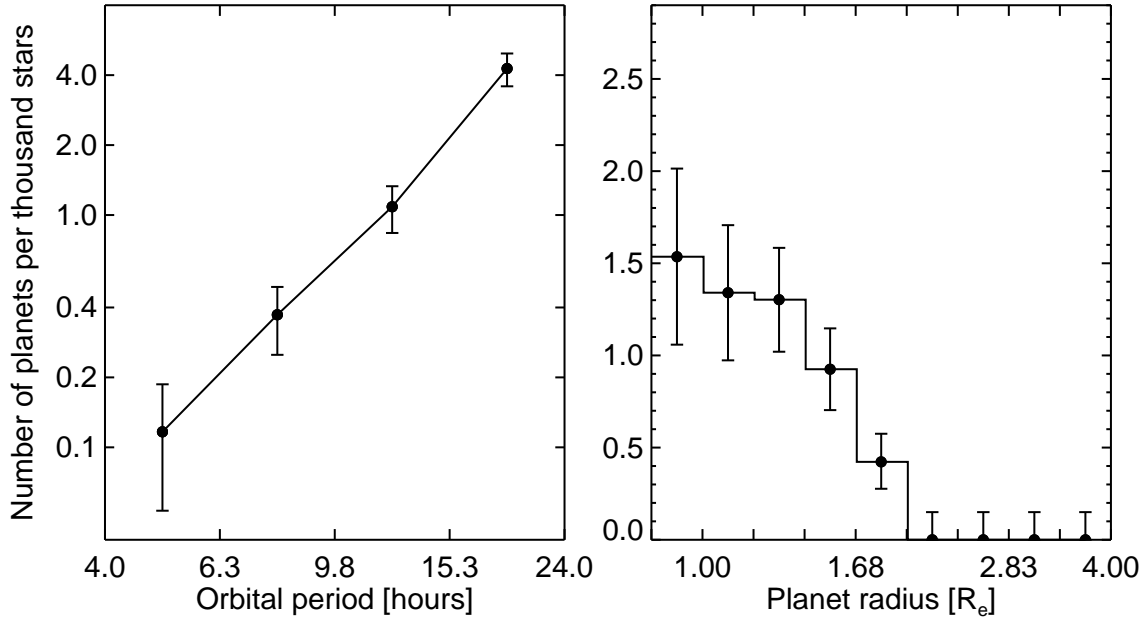}}
\end{center}
\vspace{-0.1in}
\caption{{\bf Occurrence rate for ultra-short-period planets.}
Occurrence rates were computed on a two-dimensional
grid of planet radius and orbital period.
The occurrence rate takes into account the implied
number of non-transiting planets, as well as
the number of stars that were effectively searched for
such a planet. Integrated over all radii and periods $<$1~day, the total occurrence rate is
$5.5 \pm 0.5$ planets per thousand stars.
The occurrence rate increases with orbital period, and
decreases with planet radius for $R_p<2~R_\oplus$.}
\label{fig:occur}
\vspace{0.1in}
\end{figure*}

\subsection{Method of calculation}

We have argued that the list of 106 USP candidates has a low false positive rate, and that we have a good understanding of the completeness of our pipeline for a wide range of orbital periods and planet radii. We can now use the estimate of the completeness to estimate the occurrence rate (Petigura et al. 2013). Out of the 106 USP planet candidates, 97 were detected by our automatic pipeline (see Figure~\ref{fig:diagram}; the 97 candidates belong to the box with  $\sim3,500$ candidates), although 7 of them were dropped in the subsequent visual inspection of the light curves, giving a final list of 90 USP candidates. This low fraction of dropped planets (7 out of 97) gives us confidence that our visual inspection procedure is fairly robust. In order to compare our results with the simulation, we should minimize the effects caused by this visual inspection step on the final sample of planets, since this step was not taken into account in the injection/recovery simulations.  We decided to use the 97 USP planet candidates in our group of 106 that were identified by our automatic pipeline prior to the visual inspection step, and acknowledge that a few additional planetary candidates could have been misclassified in this step.

We again concentrate on the 105,300 G and K dwarfs with $m_{\rm Kep} < 16$, the same stars used in our simulation of multiplanet systems. There are a total of 69 USP planet candidates detected by our pipeline orbiting the G and K dwarfs with orbital periods ranging from 4 to 24 hours and radii ranging from $0.84$ to 4~$R_\oplus$ (see Figure~\ref{fig:inject}). For each planet, we selected the 400 injected planets closest in orbital period and planet radius, where the distance to the $i^{\rm th}$ detected planet with coordinates $[R_i, P_i]$ is calculated with the following expression:
\begin{equation}
d^2( [R, P], [R_i, P_i]) =  \left( \frac{(P-P_i)}{P_i}\right)^2 + \left( \frac{(R-R_i)}{R_i}\right)^2 ~.
\end{equation}
This formula ensures that the positions of the 400 simulated planets are selected from a circle in log-log space. We calculated the \textit{local} completeness rate $C_i$ as the fraction of planets detected among those 400 simulated planets. The contribution of each planet to the final occurrence rate is $f_i \equiv (a/R_\star)/(C_i N_\star)$, where $N_\star$ is the total number of stars searched for planets and the factor $a/R_\star$ is the inverse transit
probability. Given a range of orbital periods and planet radii, we sum all the contributions from all $n_{\rm pl}$ planets detected in that range to estimate the final occurrence rate $f$. Following the recipe given by Howard et al.~(2012), we then define the effective number of stars searched as $n_{\rm \star, {\rm eff}} \equiv n_{\rm pl}/f$ and compute the uncertainties in $f$ from the 15.9 and 84.1 percent levels in the cumulative binomial distribution that results from drawing $n_{\rm pl}$ planets from $n_{\rm *, {\rm eff}}$ stars.  

\subsection{Dependence on planet radius and period}

Figure~\ref{fig:occur} shows the resulting occurrence rates for 4 logarithmic intervals in the orbital period (marginalized over radius) and for 9 logarithmic intervals in the planet radius (marginalized over period).  The occurrence rate increases with period, and is consistent with a power law, as had already been seen for longer orbital periods (Howard et al.\ 2012; Fressin et al.~2013; Petigura et al.~2013).  As a function of radius, the occurrence rate changes by less than a factor of $\sim 2$ in the interval 0.84--1.68~$R_\oplus$, and drops sharply for larger radii.  At ultra-short orbital periods, Earth-size planets are more common around these stars than sub-Neptunes or any other type of larger planets. Integrating over all periods and sizes, we find that there are $5.5 \pm 0.5$ planets per thousand stars (G and K dwarfs), with orbital periods in the range 4--24 hours and radii larger than 0.84~$R_\oplus$.  

\begin{figure*}[ht]
\begin{center}

\leavevmode
\hbox{
\epsfxsize=6.5in
\epsffile{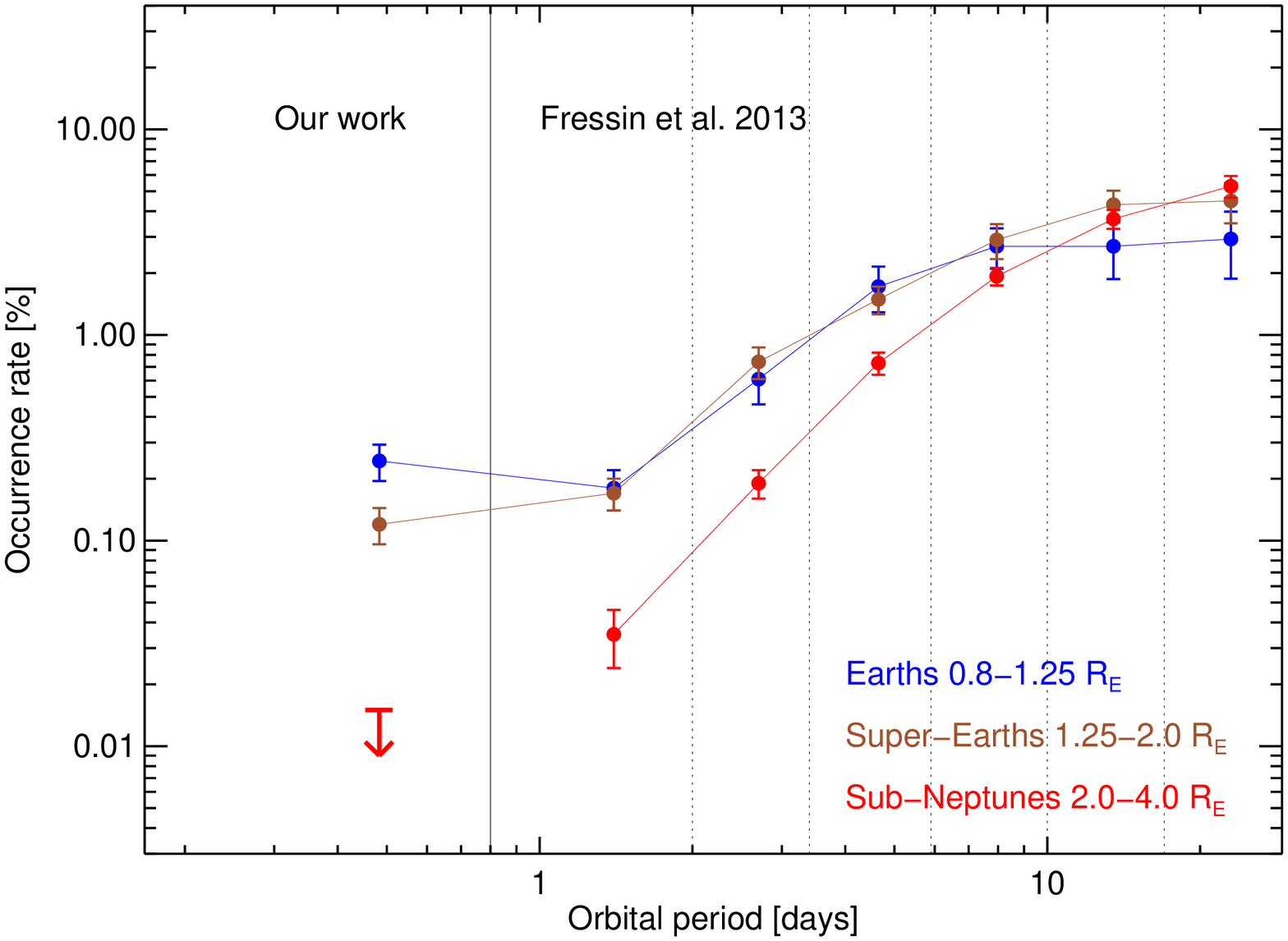}}
\end{center}
\vspace{-0.1in}
\caption{Occurrence rates for Earths, Super-Earths and Sub-Neptunes as a function of orbital period. The data points for $P>0.8$~days are from Fressin et al.\ (2013),
and the shortest-period data point is from this study.}
\label{fig:fressin}
\vspace{0.1in}
\end{figure*}

As mentioned in Sect.~\ref{sec:radius}, the lack of planets larger than 2~$R_\oplus$
may be interpreted as a consequence of the extremely strong illumination in the tight USP orbits (see, e.g., Owen \& Wu~2013 and references therein). We can compare our results to occurrence rates calculated at longer orbital periods, to try to detect trends that might support or
refute this interpretation. We use the occurrence rates measured
by Fressin et al.\ (2013) for planets as small as Earth and for periods of 0.8--85~days.
To allow a direct comparison, we divided our 50 planet candidates with orbital periods shorter than 0.8 days into three groups  according to size: Earths, super-Earths and sub-Neptunes (see Figure~\ref{fig:fressin}), and calculated the occurrence rate for each group.  The results are shown in Figure~\ref{fig:fressin}. 

A lack of sub-Neptunes is evident in Figure~\ref{fig:fressin} at the
shortest orbital periods; in fact, sub-Neptunes are significantly less common than super-Earths for orbital periods shorter than 5 days and possibly even up to 10 days.
This absence of sub-Neptunes had been predicted by different simulations of mass loss of close-in planets (Baraffe et al.\ 2004; Valencia et al.\ 2009; Lopez \& Fortney 2013a; Owen \& Wu 2013). There also seems to be a modest excess of USP Earth-size planets relative to the occurrence rate of super-Earths, which could very well be the cores of the transmuted sub-Neptunes. A more detailed statistical analysis is required to assess the significance of this excess, and it would be convenient to update the occurrence rate at longer orbital periods with new data. 

It is still possible that any mechanism that either allowed these planets to form so close to the star, or that caused these planets to have such tight orbits, is less efficient for sub-Neptunes than for smaller planets.  Obtaining the masses of some of these planets in order to measure their densities and compositions would provide some helpful context.  Obtaining precise stellar radii (via spectroscopy) and hence planet radii for a uniform sample of close-in planets, could reveal additional signatures of photoevaporation. 

\subsection{Dependence on stellar type}

We have also investigated the dependence of the occurrence rate on the type of star, to the extent that this is possible with the {\em Kepler} data.  If ultra-short period planets had an occurrence rate independent of stellar temperature, we would expect to find most planets orbiting Sun-like stars (with $T_{\rm eff} \approx 5800$K), since these stars are the most common type of star in the {\em Kepler} target catalog, and they dominate the number of USP-searchable stars.  However, the host stars of the detected USP candidates tend to be cooler, with temperatures closer to $T_{\rm eff} \approx 5000$~K (such as Kepler-78b). This might be a clue that planets with orbital periods shorter than one day are more common around cooler stars. In order to test this, we first calculate the occurrence rate for the hotter stars in our sample (G dwarfs, 5100--6100~K). Using all such stars with $m_{kep} < 16$, we have 48 USP candidates around 89,000 stars. Thus the occurrence rate for such planets is $5.1 \pm 0.7$ planets for every thousand stars. The same calculation for the cooler stars (K dwarfs, 4100--5100~K) yields $8.3 \pm 1.8$ planets per thousand stars, based on 21 USP candidates orbiting 16,200 stars.  The difference between the occurrence rates for these two spectral classes is only marginally significant (at the 1.7$\sigma$ level).  

It would be interesting if USP planets are more common around K dwarfs than G dwarfs. A similar temperature trend in the occurrence rate of planets with 1--50~day periods was found by Howard et al.\ (2012), although this finding has been challenged by Fressin et al.\ (2013) based on the statistics of planets with even longer orbital periods. This collection of results suggests that the dependence of occurrence rate on the type of star is most pronounced for the shortest period planets. Indeed, since planet formation depends on the temperature of the materials in the protoplanetary disk, and since the protoplanetary disks around cooler stars are cooler at a given orbital distance, one could imagine scenarios in which cooler stars have a higher abundance of short-period planets. In this case it will be worth extending our study to slightly longer orbital periods to find the transition point at which K and G dwarf stars have the same number of planets, and also to extend this study to include stars of spectral types F and M.

\begin{deluxetable}{lcc}
\tabletypesize{\scriptsize}
\tablecaption{Occurrence rate of USP planets for different stellar types\label{tbl:teffoccur}}
\tablewidth{0pt}

\tablehead{
\colhead{Stellar type} & \colhead{Effective temperature range}  & \colhead{Occurrence rate}
}

\startdata
F dwarfs & 6100-7100K & $0.15 \pm 0.05\%$ \\
G dwarfs & 5100-6100K & $0.51 \pm 0.07\%$ \\
K dwarfs & 4100-5100K & $0.83 \pm 0.18\%$ \\
M dwarfs & 3100-4100K & $1.10 \pm 0.40\%$ 
\enddata

\end{deluxetable}

Even though our sample of USP planets transiting F and M dwarfs is somewhat limited, we can use it to put very useful bounds on the occurrence rate of such planets. For that we repeated the completeness calculation using F and M hosts, and repeated all the steps of the computation. For F dwarfs (6100--7100~K), we find 9 planet candidates within a sample of 48,000 stars, giving an occurance rate of $1.5 \pm 0.5$ planets per thousand stars. For M dwarfs (3100--4100~K), we find 6 planet candidates out of a sample of 3537 stars, yielding a rate of $11 \pm 4$ planets per thousand stars. This represents further evidence that cooler stars are more likely to host USP planets.  If we now combine the USP planet occurrence rates for all four spectral classes of host star (M, K, G, and F) the decreasing trend in occurrence rate with increasing $T_{\rm eff}$ becomes quite significant. These results are summarized on Table~\ref{tbl:teffoccur}.

\section{Summary and Conclusions}
\label{sec:disc}

In this work we have performed a systematic search of the entire {\em Kepler} database for ultra-short period (USP) planet candidates, defined as those having orbital periods $< 1$~day.  We utilized a standard Fourier transform algorithm to search for periodic signals in the data, and found it to be quite efficient at finding short-period periodicities. An automated pipeline selected several thousand objects for further investigation including the analysis of transit-profile shapes in the folded light curves.  The folded light curves for these candidates were also inspected by eye to yield a first-cut set of 375 interesting candidates.  These objects were combined with 127 USP planet candidates from the KOI list, as well as other objects found in the literature, resulting in a set of 471 distinct candidates worthy of detailed study. 

These 471 initially selected candidates were then subjected to a number of standard tests, including examination of shifts in the light centroid during transits/eclipses, symmetry between odd and even transits/eclipses, shape of transits, etc. The final result is a set of 106 USP planet candidates that have passed a set of very restrictive tests.  Eight of these objects are completely new, while another 10 were KOIs that had been rejected, largely because their orbital period had been incorrectly identified by the {\em Kepler} pipeline.

Our final set of 106 USP candidates, and their properties, are summarized in Table 1.  In the process we also eliminated some 26 USP candidates from the KOI list and others that were found in the literature. These are listed, along with reasons for rejection, and 8 more USP candidates not considered in this study due to low SNR, in Table 2.

The USP planets are inferred to occur around one out of every 200 stars, on average.  This makes them nearly as abundant as hot Jupiters.  We also infer that the USPs nearly always have companion planets with $P<50$~days unlike hot Jupiters, which rarely have such companions. The occurrence rate of USP planets rises with period from 0.2 to 1~day, and there is evidence that the occurrence rate is higher for cooler stars than for hotter stars. The population of USP planets offers a number of opportunities for follow-up ground-based observations, as has already been illustrated by the examples of Kepler-10b (Batalha et al.\ 2011) and Kepler-78b (Sanchis-Ojeda et al.\ 2013a).

A key finding was the relative scarcity of USP planets with radius $>2~R_\oplus$. It is worth noting that the well-known USP planet 55~Cnc~e has a radius very near the top end of the range of planet sizes in our catalog; its radius has been estimated as 2.0~$R_\oplus$ (Winn et al.\ 2011) or 2.2~$R_\oplus$ (Gillon et al.\ 2012). The results of our survey would seem to imply that the discovery of an USP planet as large as 55~Cnc~e was unlikely. However, it is difficult to assess the significance of this ``fluke'' given that the mass and radius of 55~Cnc~e were determined after a process of discovery with important and complicated selection effects.

The relative scarcity of planets with $>2~R_\oplus$ could be naturally interpreted as a consequence of the strong illumination in the tight orbits. It is possible that a large fraction of the Earth-sized planets in our sample were formerly sub-Neptunes (see Owen \& Wu~2013 and references therein). There might be other observational signatures of this phenomenon, such as enhanced densities or other compositional properties, that are worth exploring. It is also worth continuing the exploration of planets with slightly longer orbital periods to determine at what distance sub-Neptunes start to become common (see Figure~\ref{fig:fressin}), and to study those systems in detail to understand the speed and efficiency of the mass-loss mechanism.

It is unclear how the USP planets attained such tight orbits, although there is little doubt that they formed further away from their host stars. The relation between the USPs and the first discovered family of close-in planets---the hot Jupiters---is also not clear.  For hot Jupiters, the formation problem is more difficult, in a sense, because they are supposed to have migrated from beyond the snow line, whereas current planet formation theories can accommodate the formation of smaller planets closer to the star.  A full comparison between the properties of both families of planets could reveal more differences that might help us understand how the close-in small planets evolve into their current stage. In particular, it would be interesting to test whether the host stars of USP planets are preferentially metal-rich, as is well known to be the case for hot Jupiters (Santos et al.\ 2004). Studies of small planets at somewhat longer periods have not found such a metallicity effect in systems with G and K host stars (Schlaufman \& Laughlin 2011; Buchhave et al.\ 2012).  It would also be interesting to measure the obliquities of the host stars to see if their rotation axes are frequently misaligned with the planetary orbits, as is the case with hot Jupiters (see, e.g., Winn et al.\ 2010; Triaud et al.\ 2010; Albrecht et al.\ 2012), or whether they have low obliquities similar to many of the multi-transit host stars that have been measured (Sanchis-Ojeda et al.\ 2012; Hirano et al.\ 2012b; Albrecht et al.\ 2013; Van Eylen et al.\ 2014). Such measurements might be challenging for small planets, but could be achievable with techniques that do not depend critically on transit observations, such as asteroseismology (Chaplin et al.\ 2013). Given a large sample of obserations, the $v \sin{i}$ technique can also be used to constrain the statistical properties of the distribution of obliquities for a given family of planets (Schlaufman 2010; Hirano et al.\ 2012a; Hirano et al.\ 2014). 

A large fraction of our planet candidates should induce radial velocity changes in their host stars at levels of a few meters per second. Measuring the masses of these planets, or constraining them, may be achievable with high-precision radial velocity instruments on large telescopes, at least for the brightest host stars.  This would increase our knowledge of the compositions of Earth-size planets.

\acknowledgements We thank Simon Albrecht, Dave Charbonneau, Brice Demory, Andrew Howard, Brian Jackson, Michael Liu, Eric Lopez, Kevin Schlaufman, Peter Sullivan, Amaury Triaud and Vincent Van Eylen for helpful discussions about short-period planets. We are also grateful to the entire {\it Kepler} team for making this study possible. R.S.O.\ and J.N.W.\ acknowledge NASA support through the Kepler Participating Scientist Program. I.E. acknowledges financial support from the ENS of Cachan. This research has made use of the NASA Exoplanet Archive, which is operated by the California Institute of Technology, under contract with the National Aeronautics and Space Administration under the Exoplanet Exploration Program. The data presented in this article were obtained from the Mikulski Archive for Space Telescopes (MAST). STScI is operated by the Association of Universities for Research in Astronomy, Inc., under NASA contract NAS5-26555. Support for MAST for non-HST data is provided by the NASA Office of Space Science via grant NNX09AF08G and by other grants and contracts.  We made use of J-band images that were obtained with the United Kingdom Infrared Telescope (UKIRT) which is operated by the Joint Astronomy Centre on behalf of the Science and Technology Facilities Council of the U.K.

\renewcommand{\arraystretch}{1.3}

\LongTables

\clearpage

\begin{deluxetable*}{rccccccccccc}
\tabletypesize{\scriptsize}
\tablecaption{Characteristics of the 106 USP planet candidates discoreved in the Kepler data  \label{tab:discovtab}}
\tablewidth{0pt}

\tablehead{
\colhead{KIC \#} & \colhead{KOI \#} &  \colhead{$m_{\rm Kep}$} & \colhead{$T_{eff} [K]$} & \colhead{$Logg$} & \colhead{$R_{\star}$ [$R_\sun$]} & \colhead{$P_{\rm orb}$ [days]} & \colhead{$t_0$ [BJD-2454900]} &\colhead{Depth [ppm]} & \colhead{Duration [hr]} &   \colhead{$a/R_{\star}$} & \colhead{$R_p$ [$R_\oplus$]} 
}

\startdata
  2711597 &    4746.01 &    14.5 &          $4861^{+ 150}_{-129}$ &  $ 4.57^{+ 0.05}_{-0.05}$ &  $ 0.72^{+ 0.08}_{-0.07}$ &   0.49020979 &      64.289540 &   $  92^{+   6}_{  -6}$ &     $ 0.910^{+ 0.055}_{-0.055}$ &  $ 3.63^{+ 0.48}_{-0.96}$ &  $ 0.76^{+ 0.12}_{-0.11}$ \\
  2718885 &       0.00 &    14.7 &          $5597^{+ 160}_{-142}$ &  $ 4.55^{+ 0.03}_{-0.28}$ &  $ 0.84^{+ 0.35}_{-0.07}$ &   0.19733350 &      64.716193 &   $  96^{+  14}_{ -14}$ &     $ 0.372^{+ 0.058}_{-0.029}$ &  $ 3.43^{+ 0.66}_{-0.94}$ &  $ 0.94^{+ 0.34}_{-0.17}$ \\
  3112129 &    4144.01 &    14.4 &          $6096^{+ 144}_{-201}$ &  $ 4.46^{+ 0.05}_{-0.31}$ &  $ 1.02^{+ 0.48}_{-0.09}$ &   0.48765719 &      64.677406 &   $ 107^{+   6}_{  -6}$ &     $ 0.899^{+ 0.045}_{-0.048}$ &  $ 3.67^{+ 0.45}_{-0.97}$ &  $ 1.21^{+ 0.50}_{-0.21}$ \\
  4665571 &    2393.00 &    14.9 &          $4894^{+ 149}_{-128}$ &  $ 4.60^{+ 0.04}_{-0.06}$ &  $ 0.70^{+ 0.09}_{-0.06}$ &   0.76669043 &      64.430840 &   $ 259^{+  13}_{ -12}$ &     $ 0.983^{+ 0.046}_{-0.046}$ &  $ 5.28^{+ 0.64}_{-1.40}$ &  $ 1.24^{+ 0.19}_{-0.17}$ \\
  4929299 &    4018.01 &    14.8 &          $5761^{+ 191}_{-164}$ &  $ 4.33^{+ 0.17}_{-0.23}$ &  $ 1.05^{+ 0.38}_{-0.19}$ &   0.43436157 &     285.738945 &   $ 192^{+   8}_{  -8}$ &     $ 1.076^{+ 0.045}_{-0.044}$ &  $ 2.74^{+ 0.32}_{-0.73}$ &  $ 1.60^{+ 0.56}_{-0.35}$ \\
  5480884 &    4841.01 &    16.0 &          $4607^{+ 123}_{-132}$ &  $ 4.73^{+ 0.06}_{-0.03}$ &  $ 0.52^{+ 0.03}_{-0.04}$ &   0.23624272 &     102.883560 &   $ 261^{+  15}_{ -16}$ &     $ 0.844^{+ 0.049}_{-0.046}$ &  $ 1.89^{+ 0.24}_{-0.50}$ &  $ 0.90^{+ 0.11}_{-0.11}$ \\
  5955905 &       0.00 &    15.0 &          $6762^{+ 183}_{-283}$ &  $ 4.26^{+ 0.11}_{-0.32}$ &  $ 1.38^{+ 0.79}_{-0.25}$ &   0.51705775 &    1207.057671 &   $ 978^{+  24}_{ -24}$ &     $ 0.753^{+ 0.019}_{-0.018}$ &  $ 4.67^{+ 0.51}_{-1.25}$ &  $ 4.84^{+ 2.54}_{-1.13}$ \\
  6359893 &       0.00 &    15.9 &          $6197^{+ 183}_{-226}$ &  $ 4.45^{+ 0.05}_{-0.30}$ &  $ 1.03^{+ 0.47}_{-0.10}$ &   0.18043173 &     285.389235 &   $ 258^{+  23}_{ -23}$ &     $ 0.385^{+ 0.043}_{-0.032}$ &  $ 3.09^{+ 0.53}_{-0.81}$ &  $ 1.90^{+ 0.76}_{-0.34}$ \\
  6525946 &    2093.03 &    15.4 &          $6487^{+ 187}_{-257}$ &  $ 4.40^{+ 0.06}_{-0.31}$ &  $ 1.12^{+ 0.56}_{-0.12}$ &   0.49607106 &     285.534133 &   $ 124^{+   7}_{  -7}$ &     $ 1.424^{+ 0.059}_{-0.059}$ &  $ 2.36^{+ 0.27}_{-0.63}$ &  $ 1.42^{+ 0.63}_{-0.26}$ \\
  8435766 &       0.00 &    11.6 &          $5085^{+ 106}_{ -93}$ &  $ 4.60^{+ 0.02}_{-0.06}$ &  $ 0.74^{+ 0.06}_{-0.03}$ &   0.35500753 &      53.604848 &   $ 229^{+   1}_{  -1}$ &     $ 0.719^{+ 0.004}_{-0.004}$ &  $ 3.37^{+ 0.37}_{-0.90}$ &  $ 1.23^{+ 0.14}_{-0.14}$ \\
  9642018 &    4430.01 &    15.5 &          $5275^{+ 169}_{-142}$ &  $ 4.59^{+ 0.03}_{-0.16}$ &  $ 0.79^{+ 0.19}_{-0.06}$ &   0.25255594 &      64.516303 &   $ 223^{+  23}_{ -20}$ &     $ 0.438^{+ 0.074}_{-0.051}$ &  $ 3.74^{+ 0.81}_{-0.96}$ &  $ 1.32^{+ 0.30}_{-0.20}$ \\
  9825174 &    2880.01 &    15.9 &          $5618^{+ 164}_{-144}$ &  $ 4.49^{+ 0.06}_{-0.27}$ &  $ 0.87^{+ 0.36}_{-0.09}$ &   0.74092963 &     102.695775 &   $ 210^{+  15}_{ -14}$ &     $ 0.894^{+ 0.069}_{-0.065}$ &  $ 5.56^{+ 0.80}_{-1.46}$ &  $ 1.44^{+ 0.53}_{-0.25}$ \\
 10006641 &    4469.01 &    14.3 &          $4981^{+ 156}_{-127}$ &  $ 4.60^{+ 0.03}_{-0.08}$ &  $ 0.73^{+ 0.11}_{-0.06}$ &   0.44708779 &      64.506995 &   $  76^{+   5}_{  -5}$ &     $ 0.773^{+ 0.051}_{-0.050}$ &  $ 3.89^{+ 0.53}_{-1.02}$ &  $ 0.70^{+ 0.12}_{-0.10}$ \\
 10527135 &    2622.01 &    15.3 &          $5430^{+ 174}_{-145}$ &  $ 4.54^{+ 0.03}_{-0.25}$ &  $ 0.84^{+ 0.32}_{-0.07}$ &   0.32868757 &      64.454984 &   $ 165^{+   9}_{  -9}$ &     $ 0.767^{+ 0.041}_{-0.043}$ &  $ 2.90^{+ 0.37}_{-0.76}$ &  $ 1.23^{+ 0.41}_{-0.20}$ \\
 10585738 &    3032.01 &    15.7 &          $5343^{+ 176}_{-156}$ &  $ 4.54^{+ 0.03}_{-0.23}$ &  $ 0.89^{+ 0.28}_{-0.06}$ &   0.63642490 &      64.688541 &   $ 196^{+  18}_{ -17}$ &     $ 0.947^{+ 0.079}_{-0.082}$ &  $ 4.51^{+ 0.70}_{-1.17}$ &  $ 1.41^{+ 0.39}_{-0.22}$ \\
 11187332 &       0.00 &    15.2 &          $5790^{+ 160}_{-172}$ &  $ 4.53^{+ 0.03}_{-0.28}$ &  $ 0.90^{+ 0.35}_{-0.07}$ &   0.30598391 &      64.826889 &   $ 126^{+  13}_{ -14}$ &     $ 0.369^{+ 0.045}_{-0.027}$ &  $ 5.42^{+ 0.94}_{-1.43}$ &  $ 1.15^{+ 0.39}_{-0.19}$ \\
 11453930 &       0.00 &    13.1 &          $6823^{+ 165}_{-242}$ &  $ 4.21^{+ 0.12}_{-0.34}$ &  $ 1.49^{+ 0.92}_{-0.29}$ &   0.22924638 &      64.512355 &   $  37^{+   3}_{  -4}$ &     $ 0.367^{+ 0.046}_{-0.025}$ &  $ 4.08^{+ 0.71}_{-1.07}$ &  $ 1.03^{+ 0.58}_{-0.26}$ \\
 11550689 &       0.00 &    14.6 &          $4168^{+ 124}_{-135}$ &  $ 4.65^{+ 0.06}_{-0.03}$ &  $ 0.60^{+ 0.05}_{-0.07}$ &   0.30160087 &      64.744663 &   $ 412^{+   6}_{  -6}$ &     $ 0.806^{+ 0.014}_{-0.014}$ &  $ 2.55^{+ 0.27}_{-0.68}$ &  $ 1.31^{+ 0.18}_{-0.19}$ \\
\hline \hline
  1717722 &    3145.02 &    15.7 &          $4812^{+ 145}_{-133}$ &  $ 4.61^{+ 0.03}_{-0.06}$ &  $ 0.71^{+ 0.07}_{-0.06}$ &   0.97730809 &     102.820321 &   $ 266^{+  21}_{ -22}$ &     $ 1.249^{+ 0.109}_{-0.095}$ &  $ 5.22^{+ 0.81}_{-1.36}$ &  $ 1.27^{+ 0.19}_{-0.18}$ \\
  3444588 &    1202.01 &    15.9 &          $4894^{+ 607}_{-904}$ &  $ 4.60^{+ 0.10}_{-0.10}$ &  $ 0.72^{+ 0.11}_{-0.11}$ &   0.92831093 &      64.945619 &   $ 397^{+  21}_{ -21}$ &     $ 1.104^{+ 0.055}_{-0.054}$ &  $ 5.70^{+ 0.69}_{-1.51}$ &  $ 1.55^{+ 0.30}_{-0.28}$ \\
  4055304 &    2119.01 &    14.1 &          $5203^{+ 171}_{-134}$ &  $ 4.51^{+ 0.05}_{-0.19}$ &  $ 0.86^{+ 0.24}_{-0.07}$ &   0.57103885 &      64.653190 &   $ 243^{+   5}_{  -5}$ &     $ 0.978^{+ 0.019}_{-0.020}$ &  $ 3.98^{+ 0.43}_{-1.07}$ &  $ 1.52^{+ 0.39}_{-0.23}$ \\
  4144576 &    2202.01 &    14.1 &          $5285^{+ 169}_{-138}$ &  $ 4.59^{+ 0.03}_{-0.16}$ &  $ 0.78^{+ 0.20}_{-0.06}$ &   0.81316598 &      64.826536 &   $ 162^{+   6}_{  -5}$ &     $ 1.237^{+ 0.045}_{-0.044}$ &  $ 4.47^{+ 0.50}_{-1.19}$ &  $ 1.12^{+ 0.26}_{-0.16}$ \\
  4852528 &     500.05 &    14.8 &          $4040^{+  64}_{-175}$ &  $ 4.70^{+ 0.07}_{-0.07}$ &  $ 0.57^{+ 0.07}_{-0.07}$ &   0.98678600 &      64.196497 &   $ 295^{+  11}_{ -11}$ &     $ 1.238^{+ 0.041}_{-0.041}$ &  $ 5.42^{+ 0.60}_{-1.44}$ &  $ 1.06^{+ 0.18}_{-0.17}$ \\
  5040077 &    3065.01 &    14.6 &          $5837^{+ 192}_{-169}$ &  $ 4.50^{+ 0.07}_{-0.28}$ &  $ 0.84^{+ 0.37}_{-0.09}$ &   0.89638348 &      65.296606 &   $ 133^{+  15}_{ -14}$ &     $ 0.965^{+ 0.098}_{-0.094}$ &  $ 6.19^{+ 1.07}_{-1.60}$ &  $ 1.11^{+ 0.43}_{-0.21}$ \\
  5095635 &    2607.01 &    14.5 &          $5883^{+ 160}_{-165}$ &  $ 4.53^{+ 0.03}_{-0.27}$ &  $ 0.89^{+ 0.33}_{-0.07}$ &   0.75445863 &      64.210363 &   $ 237^{+  16}_{ -12}$ &     $ 0.561^{+ 0.037}_{-0.047}$ &  $ 9.11^{+ 1.28}_{-2.38}$ &  $ 1.57^{+ 0.50}_{-0.25}$ \\
  5175986 &    2708.01 &    15.9 &          $4790^{+ 159}_{-132}$ &  $ 4.60^{+ 0.03}_{-0.06}$ &  $ 0.73^{+ 0.07}_{-0.05}$ &   0.86838702 &     103.275858 &   $ 511^{+  19}_{ -18}$ &     $ 0.923^{+ 0.033}_{-0.034}$ &  $ 6.40^{+ 0.72}_{-1.70}$ &  $ 1.80^{+ 0.25}_{-0.24}$ \\
  5340878 &    4199.01 &    14.3 &          $5166^{+ 148}_{-136}$ &  $ 4.65^{+ 0.03}_{-0.10}$ &  $ 0.68^{+ 0.11}_{-0.05}$ &   0.53991700 &      64.548229 &   $  96^{+   6}_{  -6}$ &     $ 1.040^{+ 0.057}_{-0.059}$ &  $ 3.51^{+ 0.45}_{-0.93}$ &  $ 0.74^{+ 0.12}_{-0.10}$ \\
  5513012 &    2668.01 &    14.2 &          $5596^{+ 155}_{-144}$ &  $ 4.59^{+ 0.03}_{-0.25}$ &  $ 0.78^{+ 0.31}_{-0.06}$ &   0.67933587 &      65.178693 &   $ 246^{+   6}_{  -7}$ &     $ 1.066^{+ 0.030}_{-0.028}$ &  $ 4.33^{+ 0.48}_{-1.16}$ &  $ 1.39^{+ 0.48}_{-0.22}$ \\
  5642620 &    2882.02 &    15.4 &          $4467^{+ 165}_{-160}$ &  $ 4.67^{+ 0.04}_{-0.06}$ &  $ 0.61^{+ 0.07}_{-0.05}$ &   0.49339463 &     285.612655 &   $ 270^{+  16}_{ -15}$ &     $ 0.870^{+ 0.052}_{-0.051}$ &  $ 3.82^{+ 0.50}_{-1.01}$ &  $ 1.10^{+ 0.17}_{-0.15}$ \\
  5942808 &    2250.02 &    15.6 &          $4922^{+ 195}_{-151}$ &  $ 4.50^{+ 0.07}_{-0.65}$ &  $ 0.82^{+ 0.69}_{-0.07}$ &   0.62628076 &     285.358414 &   $ 418^{+  15}_{ -16}$ &     $ 1.011^{+ 0.038}_{-0.036}$ &  $ 4.20^{+ 0.48}_{-1.12}$ &  $ 1.96^{+ 1.41}_{-0.36}$ \\
  5972334 &     191.03 &    15.0 &          $5700^{+ 167}_{-154}$ &  $ 4.51^{+ 0.05}_{-0.28}$ &  $ 0.87^{+ 0.37}_{-0.08}$ &   0.70862545 &      64.944413 &   $ 180^{+   8}_{  -9}$ &     $ 1.366^{+ 0.056}_{-0.057}$ &  $ 3.52^{+ 0.41}_{-0.94}$ &  $ 1.33^{+ 0.49}_{-0.23}$ \\
  6129524 &    2886.01 &    15.9 &          $5260^{+ 213}_{-163}$ &  $ 4.59^{+ 0.03}_{-0.17}$ &  $ 0.78^{+ 0.22}_{-0.06}$ &   0.88183821 &     285.152838 &   $ 268^{+  17}_{ -16}$ &     $ 1.121^{+ 0.071}_{-0.076}$ &  $ 5.31^{+ 0.72}_{-1.40}$ &  $ 1.44^{+ 0.37}_{-0.22}$ \\
  6183511 &    2542.01 &    15.5 &          $3339^{+  50}_{ -51}$ &  $ 4.96^{+ 0.06}_{-0.12}$ &  $ 0.29^{+ 0.08}_{-0.05}$ &   0.72733112 &      64.641934 &   $ 368^{+  23}_{ -22}$ &     $ 0.773^{+ 0.044}_{-0.047}$ &  $ 6.36^{+ 0.82}_{-1.67}$ &  $ 0.61^{+ 0.17}_{-0.13}$ \\
  6265792 &    2753.01 &    13.6 &          $6002^{+ 153}_{-178}$ &  $ 4.36^{+ 0.10}_{-0.27}$ &  $ 1.09^{+ 0.45}_{-0.14}$ &   0.93512027 &      65.197130 &   $  64^{+   4}_{  -4}$ &     $ 1.383^{+ 0.074}_{-0.073}$ &  $ 4.57^{+ 0.57}_{-1.21}$ &  $ 0.98^{+ 0.37}_{-0.19}$ \\
  6294819 &    2852.01 &    15.9 &          $5966^{+ 178}_{-211}$ &  $ 4.49^{+ 0.04}_{-0.30}$ &  $ 0.96^{+ 0.41}_{-0.09}$ &   0.67565200 &     285.617217 &   $ 193^{+  13}_{ -12}$ &     $ 1.702^{+ 0.085}_{-0.096}$ &  $ 2.69^{+ 0.33}_{-0.71}$ &  $ 1.52^{+ 0.56}_{-0.26}$ \\
  6310636 &    1688.01 &    14.5 &          $5738^{+ 183}_{-147}$ &  $ 4.49^{+ 0.09}_{-0.29}$ &  $ 0.82^{+ 0.38}_{-0.09}$ &   0.92103467 &      65.028092 &   $  94^{+   5}_{  -5}$ &     $ 2.328^{+ 0.076}_{-0.072}$ &  $ 2.69^{+ 0.30}_{-0.72}$ &  $ 0.90^{+ 0.37}_{-0.17}$ \\
  6362874 &    1128.01 &    13.5 &          $5487^{+ 100}_{-111}$ &  $ 4.56^{+ 0.01}_{-0.12}$ &  $ 0.83^{+ 0.12}_{-0.03}$ &   0.97486662 &      54.379852 &   $ 193^{+   2}_{  -2}$ &     $ 1.448^{+ 0.018}_{-0.018}$ &  $ 4.59^{+ 0.49}_{-1.23}$ &  $ 1.30^{+ 0.19}_{-0.16}$ \\
  6607286 &    1239.01 &    15.0 &          $6108^{+ 159}_{-234}$ &  $ 4.47^{+ 0.04}_{-0.29}$ &  $ 1.03^{+ 0.43}_{-0.10}$ &   0.78327657 &     285.814763 &   $ 264^{+   7}_{  -7}$ &     $ 1.489^{+ 0.030}_{-0.030}$ &  $ 3.58^{+ 0.38}_{-0.96}$ &  $ 1.90^{+ 0.71}_{-0.32}$ \\
  6607644 &    4159.01 &    14.5 &          $5406^{+ 170}_{-144}$ &  $ 4.58^{+ 0.02}_{-0.24}$ &  $ 0.82^{+ 0.28}_{-0.06}$ &   0.97190650 &      65.184987 &   $  64^{+   5}_{  -5}$ &     $ 1.489^{+ 0.090}_{-0.088}$ &  $ 4.40^{+ 0.58}_{-1.16}$ &  $ 0.75^{+ 0.22}_{-0.12}$ \\
  6666233 &    2306.01 &    14.8 &          $3878^{+  76}_{ -75}$ &  $ 4.73^{+ 0.06}_{-0.09}$ &  $ 0.52^{+ 0.07}_{-0.05}$ &   0.51240853 &      64.726189 &   $ 321^{+   9}_{  -9}$ &     $ 0.997^{+ 0.025}_{-0.027}$ &  $ 3.50^{+ 0.38}_{-0.94}$ &  $ 1.02^{+ 0.17}_{-0.15}$ \\
  6697756 &    2798.01 &    14.1 &          $4374^{+ 126}_{-137}$ &  $ 4.64^{+ 0.06}_{-0.03}$ &  $ 0.61^{+ 0.05}_{-0.06}$ &   0.91614360 &      65.021825 &   $  93^{+   5}_{  -5}$ &     $ 1.025^{+ 0.057}_{-0.054}$ &  $ 6.03^{+ 0.76}_{-1.59}$ &  $ 0.64^{+ 0.09}_{-0.09}$ \\
  6755944 &    4072.01 &    13.4 &          $6103^{+ 150}_{-177}$ &  $ 4.41^{+ 0.08}_{-0.27}$ &  $ 1.02^{+ 0.39}_{-0.12}$ &   0.69297791 &      53.714011 &   $  60^{+   3}_{  -3}$ &     $ 1.014^{+ 0.054}_{-0.054}$ &  $ 4.63^{+ 0.57}_{-1.22}$ &  $ 0.89^{+ 0.31}_{-0.16}$ \\
  6867588 &    2571.01 &    14.4 &          $5203^{+ 177}_{-135}$ &  $ 4.59^{+ 0.03}_{-0.14}$ &  $ 0.78^{+ 0.17}_{-0.06}$ &   0.82628363 &      64.444070 &   $ 131^{+   6}_{  -6}$ &     $ 0.866^{+ 0.049}_{-0.043}$ &  $ 6.44^{+ 0.81}_{-1.71}$ &  $ 1.00^{+ 0.21}_{-0.14}$ \\
  6934291 &    1367.01 &    15.1 &          $5076^{+ 161}_{-131}$ &  $ 4.61^{+ 0.03}_{-0.10}$ &  $ 0.73^{+ 0.12}_{-0.05}$ &   0.56785704 &      64.223769 &   $ 338^{+   8}_{  -8}$ &     $ 0.906^{+ 0.023}_{-0.023}$ &  $ 4.27^{+ 0.46}_{-1.14}$ &  $ 1.48^{+ 0.26}_{-0.20}$ \\
  6964929 &    2756.01 &    14.8 &          $5957^{+ 154}_{-187}$ &  $ 4.51^{+ 0.04}_{-0.27}$ &  $ 0.94^{+ 0.34}_{-0.08}$ &   0.66502914 &      64.655142 &   $  99^{+   5}_{  -5}$ &     $ 1.541^{+ 0.055}_{-0.052}$ &  $ 2.93^{+ 0.33}_{-0.78}$ &  $ 1.06^{+ 0.34}_{-0.17}$ \\
  6974658 &    2925.01 &    13.5 &          $5552^{+ 160}_{-135}$ &  $ 4.35^{+ 0.15}_{-0.24}$ &  $ 1.03^{+ 0.38}_{-0.17}$ &   0.71653130 &      64.560833 &   $  90^{+  17}_{ -13}$ &     $ 0.475^{+ 0.121}_{-0.096}$ &  $ 9.60^{+ 3.29}_{-2.49}$ &  $ 1.10^{+ 0.41}_{-0.25}$ \\
  7102227 &    1360.03 &    15.6 &          $5153^{+ 170}_{-143}$ &  $ 4.60^{+ 0.03}_{-0.12}$ &  $ 0.76^{+ 0.14}_{-0.06}$ &   0.76401932 &      64.458108 &   $ 131^{+  14}_{ -13}$ &     $ 0.991^{+ 0.101}_{-0.116}$ &  $ 5.16^{+ 0.93}_{-1.32}$ &  $ 0.97^{+ 0.18}_{-0.14}$ \\
  7605093 &    2817.01 &    15.8 &          $5238^{+ 202}_{-169}$ &  $ 4.60^{+ 0.03}_{-0.13}$ &  $ 0.75^{+ 0.16}_{-0.06}$ &   0.63400313 &     102.935144 &   $ 338^{+  21}_{ -21}$ &     $ 0.821^{+ 0.047}_{-0.050}$ &  $ 5.23^{+ 0.66}_{-1.38}$ &  $ 1.54^{+ 0.32}_{-0.23}$ \\
  7749002 &    4325.01 &    15.7 &          $5936^{+ 171}_{-204}$ &  $ 4.50^{+ 0.04}_{-0.31}$ &  $ 0.96^{+ 0.42}_{-0.08}$ &   0.60992303 &     102.621977 &   $ 143^{+  11}_{ -11}$ &     $ 1.118^{+ 0.084}_{-0.075}$ &  $ 3.65^{+ 0.52}_{-0.95}$ &  $ 1.32^{+ 0.49}_{-0.22}$ \\
  7826620 &    4268.01 &    15.2 &          $4294^{+ 130}_{-146}$ &  $ 4.74^{+ 0.07}_{-0.04}$ &  $ 0.50^{+ 0.04}_{-0.05}$ &   0.84990436 &      64.675047 &   $ 124^{+  11}_{ -10}$ &     $ 0.984^{+ 0.082}_{-0.087}$ &  $ 5.80^{+ 0.89}_{-1.51}$ &  $ 0.61^{+ 0.09}_{-0.09}$ \\
  7907808 &    4109.01 &    14.5 &          $4968^{+ 149}_{-126}$ &  $ 3.92^{+ 0.63}_{-0.40}$ &  $ 1.75^{+ 1.32}_{-0.97}$ &   0.65594057 &      64.444338 &   $  64^{+   4}_{  -4}$ &     $ 1.175^{+ 0.075}_{-0.073}$ &  $ 3.76^{+ 0.50}_{-0.99}$ &  $ 1.53^{+ 1.16}_{-0.86}$ \\
  8235924 &    2347.01 &    14.9 &          $3972^{+  65}_{ -60}$ &  $ 4.69^{+ 0.06}_{-0.06}$ &  $ 0.56^{+ 0.05}_{-0.05}$ &   0.58800041 &      64.471042 &   $ 300^{+   8}_{  -8}$ &     $ 0.994^{+ 0.026}_{-0.026}$ &  $ 4.03^{+ 0.44}_{-1.08}$ &  $ 1.06^{+ 0.15}_{-0.14}$ \\
  8278371 &    1150.01 &    13.3 &          $5735^{+ 107}_{-115}$ &  $ 4.34^{+ 0.11}_{-0.12}$ &  $ 1.10^{+ 0.19}_{-0.13}$ &   0.67737578 &      54.074797 &   $  71^{+   2}_{  -2}$ &     $ 1.701^{+ 0.039}_{-0.042}$ &  $ 2.71^{+ 0.29}_{-0.73}$ &  $ 1.02^{+ 0.20}_{-0.17}$ \\
  8558011 &     577.02 &    14.4 &          $5244^{+ 215}_{-169}$ &  $ 4.45^{+ 0.10}_{-0.47}$ &  $ 0.89^{+ 0.68}_{-0.11}$ &   0.63816268 &     103.039136 &   $ 111^{+   8}_{  -8}$ &     $ 1.078^{+ 0.069}_{-0.063}$ &  $ 3.98^{+ 0.53}_{-1.05}$ &  $ 1.08^{+ 0.72}_{-0.22}$ \\
  8561063 &     961.02 &    15.9 &          $3200^{+ 174}_{-174}$ &  $ 5.09^{+ 0.20}_{-0.20}$ &  $ 0.17^{+ 0.04}_{ 0.03}$ &   0.45328732 &      64.603393 &   $1885^{+  28}_{ -38}$ &     $ 0.385^{+ 0.010}_{-0.007}$ &  $ 8.01^{+ 0.87}_{-2.14}$ &  $ 0.92^{+ 0.15}_{-0.12}$ \\
  8804845 &    2039.02 &    14.3 &          $5614^{+ 147}_{-142}$ &  $ 4.56^{+ 0.03}_{-0.22}$ &  $ 0.83^{+ 0.27}_{-0.06}$ &   0.76213044 &      64.617466 &   $  63^{+   6}_{  -6}$ &     $ 1.015^{+ 0.097}_{-0.097}$ &  $ 5.01^{+ 0.84}_{-1.29}$ &  $ 0.75^{+ 0.21}_{-0.12}$ \\
  8895758 &    3106.01 &    15.4 &          $5853^{+ 177}_{-210}$ &  $ 4.46^{+ 0.05}_{-0.31}$ &  $ 1.01^{+ 0.47}_{-0.10}$ &   0.96896529 &     285.390115 &   $ 114^{+  11}_{ -12}$ &     $ 1.364^{+ 0.119}_{-0.101}$ &  $ 4.73^{+ 0.72}_{-1.23}$ &  $ 1.23^{+ 0.50}_{-0.22}$ \\
  8947520 &    2517.01 &    14.5 &          $5854^{+ 179}_{-145}$ &  $ 4.44^{+ 0.13}_{-0.30}$ &  $ 0.87^{+ 0.41}_{-0.12}$ &   0.96852469 &      64.512671 &   $ 118^{+   7}_{  -7}$ &     $ 1.155^{+ 0.060}_{-0.063}$ &  $ 5.68^{+ 0.71}_{-1.50}$ &  $ 1.06^{+ 0.46}_{-0.22}$ \\
  9092504 &    2716.01 &    15.8 &          $5693^{+ 178}_{-156}$ &  $ 4.57^{+ 0.03}_{-0.28}$ &  $ 0.80^{+ 0.34}_{-0.06}$ &   0.96286621 &     102.676246 &   $ 310^{+  13}_{ -13}$ &     $ 1.438^{+ 0.056}_{-0.055}$ &  $ 4.55^{+ 0.52}_{-1.21}$ &  $ 1.61^{+ 0.58}_{-0.26}$ \\
  9149789 &    2874.01 &    14.9 &          $5461^{+ 171}_{-137}$ &  $ 4.59^{+ 0.03}_{-0.18}$ &  $ 0.77^{+ 0.23}_{-0.06}$ &   0.35251639 &      64.663443 &   $ 163^{+   7}_{  -8}$ &     $ 0.864^{+ 0.042}_{-0.042}$ &  $ 2.76^{+ 0.33}_{-0.73}$ &  $ 1.11^{+ 0.30}_{-0.17}$ \\
  9221517 &    2281.01 &    13.8 &          $5178^{+ 172}_{-136}$ &  $ 4.56^{+ 0.03}_{-0.15}$ &  $ 0.82^{+ 0.19}_{-0.06}$ &   0.76985432 &      64.749343 &   $ 106^{+   4}_{  -4}$ &     $ 0.905^{+ 0.038}_{-0.037}$ &  $ 5.77^{+ 0.67}_{-1.53}$ &  $ 0.95^{+ 0.20}_{-0.14}$ \\
  9388479 &     936.02 &    15.1 &          $3581^{+  65}_{ -50}$ &  $ 4.81^{+ 0.08}_{-0.08}$ &  $ 0.44^{+ 0.06}_{-0.06}$ &   0.89304119 &      64.861219 &   $ 746^{+  11}_{ -11}$ &     $ 0.987^{+ 0.014}_{-0.014}$ &  $ 6.17^{+ 0.66}_{-1.65}$ &  $ 1.30^{+ 0.23}_{-0.22}$ \\
  9456281 &    4207.01 &    15.4 &          $4239^{+ 123}_{-138}$ &  $ 4.70^{+ 0.07}_{-0.04}$ &  $ 0.54^{+ 0.04}_{-0.06}$ &   0.70194345 &      65.003065 &   $ 216^{+  32}_{ -25}$ &     $ 0.526^{+ 0.099}_{-0.086}$ &  $ 8.68^{+ 2.31}_{-2.23}$ &  $ 0.85^{+ 0.13}_{-0.13}$ \\
  9472074 &    2735.01 &    15.6 &          $5154^{+ 172}_{-140}$ &  $ 4.59^{+ 0.02}_{-0.14}$ &  $ 0.80^{+ 0.14}_{-0.05}$ &   0.55884253 &      64.576962 &   $ 284^{+  17}_{ -16}$ &     $ 0.708^{+ 0.045}_{-0.047}$ &  $ 5.33^{+ 0.71}_{-1.40}$ &  $ 1.51^{+ 0.27}_{-0.20}$ \\
  9473078 &    2079.01 &    13.0 &          $5582^{+ 161}_{-153}$ &  $ 4.31^{+ 0.15}_{-0.23}$ &  $ 1.16^{+ 0.42}_{-0.20}$ &   0.69384332 &      54.078837 &   $  46^{+   2}_{  -2}$ &     $ 1.355^{+ 0.052}_{-0.051}$ &  $ 3.47^{+ 0.40}_{-0.92}$ &  $ 0.87^{+ 0.31}_{-0.19}$ \\
  9580167 &    2548.01 &    15.0 &          $4865^{+ 183}_{-160}$ &  $ 4.61^{+ 0.03}_{-0.07}$ &  $ 0.70^{+ 0.09}_{-0.06}$ &   0.82715274 &     285.928326 &   $ 270^{+  34}_{ -33}$ &     $ 1.114^{+ 0.156}_{-0.123}$ &  $ 4.87^{+ 0.98}_{-1.27}$ &  $ 1.27^{+ 0.21}_{-0.19}$ \\
  9787239 &     952.05 &    15.8 &          $3727^{+ 104}_{ -64}$ &  $ 4.76^{+ 0.08}_{-0.08}$ &  $ 0.50^{+ 0.06}_{-0.06}$ &   0.74295778 &      64.794040 &   $ 221^{+  21}_{ -21}$ &     $ 1.004^{+ 0.100}_{-0.101}$ &  $ 4.94^{+ 0.84}_{-1.27}$ &  $ 0.80^{+ 0.14}_{-0.13}$ \\
 10024051 &    2409.01 &    14.9 &          $5256^{+ 168}_{-139}$ &  $ 4.61^{+ 0.03}_{-0.12}$ &  $ 0.72^{+ 0.15}_{-0.05}$ &   0.57736948 &      64.270976 &   $ 395^{+   8}_{  -8}$ &     $ 0.997^{+ 0.019}_{-0.019}$ &  $ 3.94^{+ 0.42}_{-1.06}$ &  $ 1.60^{+ 0.33}_{-0.23}$ \\
 10028535 &    2493.01 &    15.3 &          $5166^{+ 185}_{-144}$ &  $ 4.58^{+ 0.02}_{-0.13}$ &  $ 0.79^{+ 0.15}_{-0.06}$ &   0.66308666 &      64.978343 &   $ 321^{+  12}_{ -12}$ &     $ 0.991^{+ 0.033}_{-0.034}$ &  $ 4.55^{+ 0.51}_{-1.21}$ &  $ 1.58^{+ 0.30}_{-0.22}$ \\
 10281221 &    3913.01 &    15.0 &          $6263^{+ 169}_{-211}$ &  $ 4.46^{+ 0.05}_{-0.28}$ &  $ 1.01^{+ 0.40}_{-0.10}$ &   0.58289571 &      64.620720 &   $ 581^{+  11}_{ -11}$ &     $ 0.712^{+ 0.017}_{-0.015}$ &  $ 5.57^{+ 0.60}_{-1.49}$ &  $ 2.75^{+ 0.97}_{-0.46}$ \\
 10319385 &    1169.01 &    13.2 &          $5676^{+ 100}_{-117}$ &  $ 4.53^{+ 0.02}_{-0.14}$ &  $ 0.92^{+ 0.15}_{-0.03}$ &   0.68920948 &      53.191352 &   $ 187^{+   2}_{  -2}$ &     $ 1.456^{+ 0.020}_{-0.019}$ &  $ 3.22^{+ 0.34}_{-0.86}$ &  $ 1.41^{+ 0.22}_{-0.18}$ \\
 10468885 &    2589.01 &    15.6 &          $5177^{+ 178}_{-137}$ &  $ 4.58^{+ 0.02}_{-0.15}$ &  $ 0.80^{+ 0.17}_{-0.05}$ &   0.66407444 &      65.075194 &   $ 221^{+  19}_{ -19}$ &     $ 0.958^{+ 0.078}_{-0.085}$ &  $ 4.66^{+ 0.72}_{-1.21}$ &  $ 1.34^{+ 0.27}_{-0.20}$ \\
 10604521 &    2797.01 &    15.8 &          $6173^{+ 140}_{-276}$ &  $ 4.43^{+ 0.05}_{-0.32}$ &  $ 1.11^{+ 0.59}_{-0.12}$ &   0.86812153 &     285.335843 &   $ 245^{+  13}_{ -13}$ &     $ 1.398^{+ 0.053}_{-0.051}$ &  $ 4.22^{+ 0.48}_{-1.13}$ &  $ 1.98^{+ 0.92}_{-0.36}$ \\
 10647452 &    4366.01 &    15.6 &          $5377^{+ 223}_{-170}$ &  $ 4.59^{+ 0.04}_{-0.17}$ &  $ 0.74^{+ 0.23}_{-0.06}$ &   0.76295078 &     285.410004 &   $ 197^{+  18}_{ -17}$ &     $ 0.865^{+ 0.092}_{-0.074}$ &  $ 5.83^{+ 1.00}_{-1.50}$ &  $ 1.18^{+ 0.33}_{-0.19}$ \\
 10975146 &    1300.01 &    14.3 &          $4441^{+ 142}_{-137}$ &  $ 4.71^{+ 0.05}_{-0.03}$ &  $ 0.53^{+ 0.04}_{-0.04}$ &   0.63133223 &      64.418552 &   $ 423^{+   5}_{  -5}$ &     $ 0.993^{+ 0.013}_{-0.012}$ &  $ 4.33^{+ 0.47}_{-1.16}$ &  $ 1.18^{+ 0.15}_{-0.15}$ \\
 11030475 &    2248.03 &    15.5 &          $5290^{+ 173}_{-142}$ &  $ 4.60^{+ 0.03}_{-0.15}$ &  $ 0.75^{+ 0.19}_{-0.06}$ &   0.76196246 &      64.259895 &   $ 182^{+  11}_{ -10}$ &     $ 1.123^{+ 0.059}_{-0.060}$ &  $ 4.59^{+ 0.57}_{-1.22}$ &  $ 1.13^{+ 0.26}_{-0.17}$ \\
 11197853 &    2813.01 &    13.6 &          $5143^{+ 164}_{-141}$ &  $ 4.64^{+ 0.05}_{-1.10}$ &  $ 0.62^{+ 1.88}_{-0.04}$ &   0.69846257 &      53.675039 &   $ 120^{+  10}_{  -9}$ &     $ 0.732^{+ 0.066}_{-0.066}$ &  $ 6.38^{+ 1.02}_{-1.65}$ &  $ 0.84^{+ 2.12}_{-0.18}$ \\
 11246161 &    2796.01 &    14.8 &          $6108^{+ 143}_{-196}$ &  $ 4.47^{+ 0.05}_{-0.27}$ &  $ 1.00^{+ 0.38}_{-0.09}$ &   0.53740984 &      64.495864 &   $  93^{+   6}_{  -6}$ &     $ 1.037^{+ 0.065}_{-0.062}$ &  $ 3.49^{+ 0.46}_{-0.92}$ &  $ 1.09^{+ 0.36}_{-0.18}$ \\
 11401182 &    1428.01 &    14.6 &          $4911^{+ 157}_{-128}$ &  $ 4.53^{+ 0.06}_{-0.65}$ &  $ 0.78^{+ 0.11}_{-0.07}$ &   0.92785963 &      64.122301 &   $ 492^{+   5}_{  -5}$ &     $ 1.224^{+ 0.013}_{-0.014}$ &  $ 5.17^{+ 0.55}_{-1.39}$ &  $ 1.90^{+ 0.31}_{-0.27}$ \\
 11547505 &    1655.01 &    13.8 &          $5902^{+ 155}_{-168}$ &  $ 4.44^{+ 0.07}_{-0.27}$ &  $ 0.99^{+ 0.39}_{-0.11}$ &   0.93846561 &      63.909869 &   $ 206^{+   5}_{  -5}$ &     $ 1.176^{+ 0.024}_{-0.025}$ &  $ 5.43^{+ 0.58}_{-1.46}$ &  $ 1.60^{+ 0.57}_{-0.28}$ \\
 11600889 &    1442.01 &    12.5 &          $5626^{+  92}_{-124}$ &  $ 4.40^{+ 0.06}_{-0.15}$ &  $ 1.07^{+ 0.21}_{-0.09}$ &   0.66931017 &      53.745653 &   $ 122^{+   2}_{  -2}$ &     $ 1.461^{+ 0.020}_{-0.022}$ &  $ 3.12^{+ 0.33}_{-0.84}$ &  $ 1.31^{+ 0.25}_{-0.19}$ \\
 11752632 &    2492.01 &    13.8 &          $6060^{+ 145}_{-170}$ &  $ 4.51^{+ 0.04}_{-0.28}$ &  $ 0.93^{+ 0.36}_{-0.08}$ &   0.98493890 &      64.350365 &   $  75^{+   4}_{  -4}$ &     $ 1.828^{+ 0.062}_{-0.067}$ &  $ 3.67^{+ 0.41}_{-0.98}$ &  $ 0.91^{+ 0.31}_{-0.15}$ \\
 11870545 &    1510.01 &    15.9 &          $4924^{+ 156}_{-129}$ &  $ 4.60^{+ 0.03}_{-0.07}$ &  $ 0.74^{+ 0.08}_{-0.06}$ &   0.83996160 &     102.954242 &   $ 521^{+  19}_{ -20}$ &     $ 1.118^{+ 0.042}_{-0.039}$ &  $ 5.10^{+ 0.58}_{-1.36}$ &  $ 1.84^{+ 0.27}_{-0.25}$ \\
 11904151 &      72.01 &    11.0 &          $5627^{+  44}_{ -44}$ &  $ 4.34^{+ 0.03}_{-0.03}$ &  $ 1.06^{+ 0.02}_{-0.02}$ &   0.83749060 &      53.687909 &   $ 177^{+   1}_{  -1}$ &     $ 1.634^{+ 0.010}_{-0.009}$ &  $ 3.49^{+ 0.38}_{-0.94}$ &  $ 1.54^{+ 0.15}_{-0.16}$ \\
 12170648 &    2875.01 &    14.8 &          $5180^{+ 168}_{-138}$ &  $ 4.55^{+ 0.05}_{-0.13}$ &  $ 0.78^{+ 0.16}_{-0.07}$ &   0.29969767 &      64.717608 &   $ 344^{+  11}_{ -11}$ &     $ 0.643^{+ 0.028}_{-0.028}$ &  $ 3.16^{+ 0.37}_{-0.84}$ &  $ 1.61^{+ 0.33}_{-0.24}$ \\
 12265786 &    4595.01 &    15.5 &          $4985^{+ 152}_{-135}$ &  $ 4.68^{+ 0.03}_{-0.09}$ &  $ 0.63^{+ 0.09}_{-0.04}$ &   0.59701811 &      64.713396 &   $ 268^{+  41}_{ -33}$ &     $ 0.464^{+ 0.076}_{-0.074}$ &  $ 8.41^{+ 2.06}_{-2.17}$ &  $ 1.15^{+ 0.19}_{-0.17}$ \\
 12405333 &    3009.01 &    15.2 &          $5231^{+ 183}_{-140}$ &  $ 4.48^{+ 0.08}_{-0.27}$ &  $ 0.86^{+ 0.30}_{-0.09}$ &   0.76486301 &      64.368844 &   $ 154^{+  13}_{ -13}$ &     $ 1.017^{+ 0.084}_{-0.085}$ &  $ 5.05^{+ 0.78}_{-1.31}$ &  $ 1.20^{+ 0.38}_{-0.21}$ \\
\hline \hline
  3834322 &    2763.01 &    15.4 &          $4787^{+ 144}_{-128}$ &  $ 4.61^{+ 0.03}_{-0.06}$ &  $ 0.71^{+ 0.07}_{-0.05}$ &   0.49843515 &      64.569826 &   $ 237^{+  13}_{ -12}$ &     $ 0.828^{+ 0.045}_{-0.043}$ &  $ 4.06^{+ 0.51}_{-1.07}$ &  $ 1.20^{+ 0.17}_{-0.16}$ \\
  5008501 &    1666.01 &    14.5 &          $6444^{+ 174}_{-227}$ &  $ 4.38^{+ 0.06}_{-0.30}$ &  $ 1.20^{+ 0.63}_{-0.15}$ &   0.96802576 &      64.756733 &   $ 203^{+   9}_{  -9}$ &     $ 0.707^{+ 0.034}_{-0.035}$ &  $ 9.29^{+ 1.12}_{-2.47}$ &  $ 1.94^{+ 0.91}_{-0.38}$ \\
  5080636 &    1843.00 &    14.4 &          $3584^{+  69}_{ -54}$ &  $ 4.80^{+ 0.06}_{-0.09}$ &  $ 0.45^{+ 0.08}_{-0.05}$ &   0.17689162 &      64.552254 &   $ 155^{+   8}_{  -8}$ &     $ 0.581^{+ 0.027}_{-0.030}$ &  $ 2.07^{+ 0.25}_{-0.55}$ &  $ 0.61^{+ 0.12}_{-0.10}$ \\
  5773121 &    4002.01 &    15.0 &          $5396^{+ 155}_{-133}$ &  $ 4.49^{+ 0.05}_{-0.23}$ &  $ 0.93^{+ 0.30}_{-0.08}$ &   0.52417582 &      64.601474 &   $ 240^{+  11}_{ -12}$ &     $ 0.709^{+ 0.043}_{-0.038}$ &  $ 4.98^{+ 0.65}_{-1.31}$ &  $ 1.63^{+ 0.47}_{-0.26}$ \\
  5980208 &    2742.01 &    15.0 &          $4258^{+ 120}_{-146}$ &  $ 4.63^{+ 0.06}_{-0.02}$ &  $ 0.62^{+ 0.04}_{-0.07}$ &   0.78916057 &      64.505441 &   $ 225^{+  12}_{ -11}$ &     $ 0.834^{+ 0.044}_{-0.046}$ &  $ 6.41^{+ 0.80}_{-1.69}$ &  $ 1.00^{+ 0.14}_{-0.15}$ \\
  6750902 &    3980.01 &    14.7 &          $5888^{+ 159}_{-160}$ &  $ 4.39^{+ 0.11}_{-0.25}$ &  $ 1.01^{+ 0.36}_{-0.14}$ &   0.46933249 &      64.742598 &   $ 316^{+  11}_{ -10}$ &     $ 0.743^{+ 0.029}_{-0.031}$ &  $ 4.30^{+ 0.49}_{-1.15}$ &  $ 2.00^{+ 0.67}_{-0.38}$ \\
  7051984 &    2879.01 &    12.8 &          $5653^{+ 165}_{-145}$ &  $ 3.85^{+ 0.44}_{-0.22}$ &  $ 2.18^{+ 0.87}_{-1.03}$ &   0.33906981 &      53.680375 &   $  47^{+   1}_{  -1}$ &     $ 0.987^{+ 0.033}_{-0.032}$ &  $ 2.33^{+ 0.26}_{-0.62}$ &  $ 1.61^{+ 0.70}_{-0.76}$ \\
  7269881 &    2916.01 &    14.4 &          $5088^{+ 160}_{-130}$ &  $ 4.58^{+ 0.04}_{-0.09}$ &  $ 0.74^{+ 0.12}_{-0.06}$ &   0.30693783 &      64.635907 &   $ 134^{+   9}_{  -9}$ &     $ 0.464^{+ 0.043}_{-0.036}$ &  $ 4.39^{+ 0.70}_{-1.14}$ &  $ 0.95^{+ 0.17}_{-0.14}$ \\
  7582691 &    4419.01 &    15.2 &          $3939^{+  61}_{ -51}$ &  $ 4.74^{+ 0.06}_{-0.06}$ &  $ 0.53^{+ 0.05}_{-0.05}$ &   0.25981944 &     285.437765 &   $ 281^{+  30}_{ -26}$ &     $ 0.423^{+ 0.053}_{-0.052}$ &  $ 4.05^{+ 0.81}_{-1.04}$ &  $ 0.96^{+ 0.14}_{-0.14}$ \\
  8189801 &    2480.01 &    15.7 &          $3990^{+  84}_{ -66}$ &  $ 4.69^{+ 0.06}_{-0.11}$ &  $ 0.55^{+ 0.09}_{-0.05}$ &   0.66682749 &      64.706275 &   $ 539^{+  26}_{ -25}$ &     $ 0.712^{+ 0.038}_{-0.037}$ &  $ 6.33^{+ 0.80}_{-1.67}$ &  $ 1.42^{+ 0.25}_{-0.21}$ \\
  8416523 &    4441.01 &    14.9 &          $5017^{+ 160}_{-132}$ &  $ 4.59^{+ 0.03}_{-0.09}$ &  $ 0.75^{+ 0.11}_{-0.06}$ &   0.34184200 &      64.318576 &   $ 279^{+   6}_{  -6}$ &     $ 0.856^{+ 0.019}_{-0.019}$ &  $ 2.72^{+ 0.29}_{-0.73}$ &  $ 1.39^{+ 0.22}_{-0.19}$ \\
  9353742 &    3867.01 &    14.9 &          $5825^{+ 157}_{-180}$ &  $ 4.52^{+ 0.03}_{-0.30}$ &  $ 0.92^{+ 0.38}_{-0.07}$ &   0.93875053 &      64.682273 &   $ 272^{+   7}_{  -7}$ &     $ 1.240^{+ 0.028}_{-0.027}$ &  $ 5.15^{+ 0.56}_{-1.38}$ &  $ 1.73^{+ 0.62}_{-0.28}$ \\
  9467404 &    2717.01 &    12.4 &          $6437^{+ 151}_{-205}$ &  $ 4.25^{+ 0.14}_{-0.31}$ &  $ 1.33^{+ 0.77}_{-0.26}$ &   0.92990607 &      53.162180 &   $ 105^{+   1}_{  -1}$ &     $ 1.601^{+ 0.019}_{-0.020}$ &  $ 3.96^{+ 0.42}_{-1.06}$ &  $ 1.53^{+ 0.82}_{-0.37}$ \\
  9475552 &    2694.01 &    15.0 &          $4818^{+ 142}_{-108}$ &  $ 4.48^{+ 0.09}_{-0.89}$ &  $ 0.86^{+ 1.90}_{-0.08}$ &   0.84338039 &      64.122461 &   $ 337^{+   7}_{  -7}$ &     $ 1.253^{+ 0.026}_{-0.026}$ &  $ 4.58^{+ 0.49}_{-1.22}$ &  $ 1.89^{+ 3.60}_{-0.40}$ \\
  9873254 &     717.00 &    13.4 &          $5665^{+ 105}_{-127}$ &  $ 4.39^{+ 0.07}_{-0.17}$ &  $ 1.09^{+ 0.25}_{-0.09}$ &   0.90037004 &      53.519695 &   $  34^{+   3}_{  -3}$ &     $ 1.255^{+ 0.100}_{-0.099}$ &  $ 4.81^{+ 0.71}_{-1.26}$ &  $ 0.72^{+ 0.16}_{-0.11}$ \\
  9885417 &    3246.01 &    12.4 &          $4854^{+ 130}_{-107}$ &  $ 3.94^{+ 0.62}_{-0.43}$ &  $ 1.75^{+ 1.40}_{-0.96}$ &   0.68996781 &      53.595490 &   $ 113^{+   2}_{  -2}$ &     $ 1.042^{+ 0.024}_{-0.023}$ &  $ 4.51^{+ 0.49}_{-1.21}$ &  $ 2.03^{+ 1.62}_{-1.14}$ \\
  9967771 &    1875.02 &    14.5 &          $5953^{+ 158}_{-175}$ &  $ 4.49^{+ 0.05}_{-0.28}$ &  $ 0.93^{+ 0.37}_{-0.09}$ &   0.53835457 &      64.648809 &   $ 185^{+   5}_{  -5}$ &     $ 1.344^{+ 0.029}_{-0.029}$ &  $ 2.73^{+ 0.29}_{-0.73}$ &  $ 1.43^{+ 0.51}_{-0.24}$ \\
 10186945 &    4070.01 &    14.5 &          $5080^{+ 178}_{-137}$ &  $ 4.53^{+ 0.04}_{-0.13}$ &  $ 0.83^{+ 0.16}_{-0.06}$ &   0.39683023 &      64.395900 &   $ 170^{+   7}_{  -7}$ &     $ 0.753^{+ 0.033}_{-0.033}$ &  $ 3.58^{+ 0.42}_{-0.95}$ &  $ 1.20^{+ 0.23}_{-0.17}$ \\
 12115188 &    2396.01 &    14.7 &          $5529^{+ 158}_{-142}$ &  $ 4.47^{+ 0.06}_{-0.26}$ &  $ 0.94^{+ 0.35}_{-0.09}$ &   0.49543428 &      64.477019 &   $ 281^{+   6}_{  -6}$ &     $ 1.111^{+ 0.023}_{-0.026}$ &  $ 3.04^{+ 0.33}_{-0.81}$ &  $ 1.77^{+ 0.59}_{-0.29}$ 
\enddata
\tablenotetext{a}{Stellar parameters obtained from the compilation by Huber et al.~(2014).}
\tablenotetext{b}{Transit parameters based on an analysis of the folded transit light curve.}
\tablenotetext{c}{The first 18 planet candidates are new planet candidates. These objects either
emerged from our search for the first time, or appear on the KOI list as false positives but that we consider to be viable candidates. After the double horizontal lines, the 69 KOIs that passed our false positive tests (Akeson et al.~2013). Finally,  after the next double horizontal lines, 19 other candidates that were discovered by other authors (Ofir \& Dreizler~2013, Huang et al.~2013, and Jackson et al.~2013).}
\tablenotetext{d}{Planets discovered transiting stars with other KOI planets with no KOI number assigned are labelled with the KOI of the star followed by a dot and two zeros.}
\end{deluxetable*}

\begin{center}
\begin{deluxetable*}{ccccc}
\tabletypesize{\scriptsize}
\tablecaption{USP planet candidates that appear in the literature but do not appear in our final list\tablenotemark{a} \label{tab:removed}}
\tablewidth{0pt}

\tablehead{
\colhead{SNR < 12} & \colhead{Odd-even difference} &  \colhead{Phase curve analysis} & \colhead{Centroid shift} & \colhead{Other\tablenotemark{b}} 
}

\startdata

KIC2986833 / KOI4875.01 & KIC4844004 / KOI1662.01 & KIC6690836 / KOI2699.01 & KIC2558370 / KOI3855.01 & KIC5475431 / KOI1546.01 (BD) \\
KIC3550372 / KOI4357.01 & KIC5387843 / KOI1762.01 & KIC8564672 / HUANG13 &   KIC5436161 / KOI4351.01 &KIC6526710 / OFIR13 (NF)  \\
KIC5033823 / KOI4546.01 & KIC5801571 / KOI225.01 &  KIC10453521 / JACKSON13 & KIC6185496 / HUANG13 &  KIC8639908 / KOI2700.01 (ATP) \\
KIC5208962 / KOI4912.01 & KIC6028860 / KOI2950.01 & KIC11456279 / KOI3204.01 & KIC6209798 / HUANG13 &  KIC12557548 / KOI3794.01 (ATP) \\
KIC5371777 / KOI4327.01 & KIC6129694 / KOI4131.01 & KIC12692087 / HUANG13 &  KIC6279974 / KOI1812.01 & \\
KIC7668416 / KOI3089.01 & KIC7681230 / KOI4294.01 &   &  KIC9752982 / KOI3871.01 & \\
KIC10140122 / KOI4862.01 & KIC9044228 / KOI1988.01 &  &  KIC12069414 / KOI4512.01 & \\
KIC11973517 / KOI4862.01 & KIC9535585 / HUANG13 &     &          & \\
  & KIC11774303 / KOI2269.01 &      &     &  \\        
 & KIC12109845 / KOI3765.01 &       &    & 
\enddata
\tablenotetext{a}{Those planets with no assigned KOI number have a name tag built from the first name of the first author of the discovery paper followed by the two digit year of publication (Ofir \& Dreizler~2013, Huang et al.~2013, and Jackson et al.~2013).}
\tablenotetext{b}{BD stands for likely Brown Dwarf. ATP stands for asymmetric transit profile, likely due to a dust tail emerging from the planet (Rappaport et al. 2012; Rappaport et al. 2013b). NF stands for not found in the data. }

\end{deluxetable*}
\end{center}

\end{document}